\documentclass[journal]{IEEEtran}
\ifCLASSINFOpdf
\else
\fi
\usepackage{cite}
\usepackage{graphicx}
\usepackage{amsmath,amssymb,amsfonts}
\usepackage{booktabs}%table's \toprule
\usepackage{multirow}%table for three line
\usepackage{makecell}
\usepackage[subrefformat=parens,labelformat=parens]{subfig}%multiple figures
\usepackage[ruled]{algorithm2e}
\begin{document}
\title{An Intelligent Group Event Recommendation System in Social networks}

\author{\textbf{Guoqiong~Liao$^{1,2}$,
        Xiaomei~Huang$^{1,4,*}$,
        Neal N. Xiong$^{3}$,
       and~Changxuan~Wan}$^{1,2}$\\% <-this % stops a space
\footnotesize{$^{1}$School of Information Management,  Jiangxi University of Finance and Economics,  Nanchang 330013, China.\\
$^{2}$Jiangxi Province Key Laboratory of Data and Knowledge Engineering (Jiangxi University of Finance and Economics), Nanchang 330013, China.\\
$^{3}$Department of Mathematics and Computer Science, Northeastern State University, OK 74464 USA.\\
$^{4}$School of Mathematics and Information Science, Jiangxi Normal University, Nanchang 330022, China.\\
$^{*}$Corresponding Author: Xiaomei Huang. Email: huangxm501@126.com.}
}

\maketitle

\begin{abstract}
The importance of contexts has been widely recognized in recommender systems for individuals. However, most existing group recommendation models in Event-Based Social Networks (EBSNs) focus on how to aggregate group members' preferences to form group preferences. In these models, the influence of contexts on groups is considered but simply defined in a manual way, which cannot model the complex and deep interactions between contexts and groups. In this paper, we propose an Attention-based Context-aware Group Event Recommendation model (ACGER) in EBSNs. ACGER models the deep, non-linear influence of contexts on users, groups, and events through multi-layer neural networks. Especially, a novel attention mechanism is designed to enable the influence weights of contexts on users/groups change dynamically with the events concerned. Considering that groups may have completely different behavior patterns from group members, we propose that the preference of a group need to be obtained from indirect and direct perspectives (called indirect preference and direct preference respectively). In order to obtain the indirect preference, we propose a method of aggregating preferences based on attention mechanism. Compared with existing predefined strategies, this method can flexibly adapt the strategy according to the events concerned by the group. In order to obtain the  direct preference, we employ neural networks to directly learn it from group-event interactions. Furthermore, to make full use of rich user-event interactions in EBSNs, we integrate the context-aware individual recommendation task into ACGER , which enhances the accuracy of learning of user embeddings and event embeddings. Extensive experiments on two real datasets from Meetup show that our model ACGER significantly outperforms the state-of-the-art models.
\end{abstract}


% \fi
\IEEEpeerreviewmaketitle

\section{Introduction}
\IEEEPARstart{E}{vent-based} Social Networks (EBSNs) applications, such as Meetup.com, Douban.com, and Plancast.com, have become increasing popular in recent years. EBSNs provide online platforms for users to create, distribute, organize and register all kinds of social events, which promotes the success of offline interactions among users. The events here could be academic meetings, business exhibitions, dining out, and movies night, etc. In order to alleviate the problem of information overload brought by massive events, many models on recommending events for individuals are proposed \cite{zhang2015collective,wang2018adapting,liao2018global,qiao2014event}.

Since people often participate in offline events in groups in real life, recommending  events for groups of people has become research focus in recent years \cite{boratto2015art, ricci2015recommender, wang2016member, Seo2018Enhanced}. Traditional group recommendation methods focus on how to aggregate member preferences to form the group preference, and pay less attention to the influence of contexts (such as time, location, and social relationship) on group preferences. In fact, contexts may have important impacts on group behaviors. For example, when a group decides on which restaurant is suitable for dinning out, besides the type of food, the contextual factors such as the restaurant's location, the parking lot's capacity, and the members' free time will all have impacts on the group's decision.

The impacts of contexts have been widely studied in event recommender systems for individuals \cite{zhang2015collective, qiao2014event, macedo2015context, xu2019semantic}. These works study the influence mechanism of contexts on individuals by defining linear or non-linear functions manually. However, manually defined functions are not sufficient to capture the deep, highly non-liner interactions among entities such as users, events, and contexts. Moreover, these works focus on the impacts of contexts on individuals rather than groups, so they cannot be directly applied to event recommendation for groups.

Recently Du et al. \cite{Du2020GERF} studied the group event recommendation problem considering influences of contexts including event content, time, location, and social relationship, and proposed a group recommendation method based on learning-to-rank technology. This method incorporates the influences of various contexts, but the influence functions are defined manually. And the method does not differentiate the influence weights of various contexts. There is a recent work modeling the contextual influences in the field of context-aware recommendation for individuals \cite{Mei2018Attentive}. It proposes a neural attention mechanism to model the influence weights of contexts on users and events. However, its measurement of influence weights of contexts on users only considers the interaction between contexts and users, neglecting the impacts of events, which is not quite consistent with the fact. For example, the influence weight of the context of season on users will change with the type of events. When a user is faced with winter skiing events, the influence weight of the season is great, but when the user is faced with movie events, the influence weight of the season becomes smaller.

\begin{figure}[h]
	\centering
	\includegraphics[width=.8\linewidth]{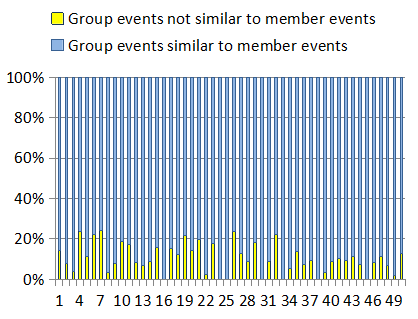}
	\caption{Characteristics of group events attended by 50 groups in New York city in 2016 on Meetup website.}
	\label{fig:f1}
\end{figure}

How to acquire group preferences accurately is the core problem of group recommendation.  Existing group recommendation algorithms focus on obtaining group preferences by aggregating members' preferences, and various group aggregation strategies have been proposed, such as strategies based on social theory (like approval voting, least misery, average, etc. \cite{Seo2018Enhanced, Masthoff2004Group}) and strategies considering users' special needs or users' expertise \cite{ardissono2003intrigue, berkovsky2010group,Quintarelli2016Recommending}.  However, these works exist following two limitations: (1) predefined group aggregation strategies lack flexibility. When the type of events concerned  changes, the group may adopt a different decision-making strategy, and the predefined strategy cannot adapt to it. For example, a tour group often goes hiking in various scenic spots, and the group usually adopts the average strategy to decide the next spot (i.e., each member has equal weight to the final group decision). When the group is faced with some different type of tour project (such as surfing or skiing), members with experiences in such project may have greater weights to the final decision. Then the previous average strategy is no longer applicable. Therefore, we need to study a more flexible and adaptive group aggregation strategy.(2) a group preference is not completely determined by the historical preferences of group members. For example, in real life, an individual user could choose jogging or walking as his/her leisure sport. When he/she is in a group, the group may still choose jogging or walking as its leisure sport, but may also choose playing football, basketball or other sports which need cooperation with each other. We analyzed the events attended by 50 groups of New York City in 2016 on Meetup website. Fig. \ref{fig:f1} shows the characteristics of these group events. It can be seen that a large proportion of group events in most groups are similar to the historical events of members, but there are also proportions of group events that are not similar to the historical events of any member. This suggests that group behaviors may be completely different from individual behaviors. For the convenience of discussion, we call the group preference obtained from the historical preferences of group members as group indirect preference, and call the group preference completely different from the historical preferences of members as group direct preference. Therefore, a reasonable group recommendation model should model both indirect preference and direct preference of a group.

To address above problems, we propose an Attention-based Context-aware Group Event Recommendation model (ACGER) for EBSNs. The details are as follows.

In order to characterize the complex influence of contexts on users, groups and events, we propose a deep model to learn the representations of users, groups, and events under the influence of contexts. This is motivated by the successful development of deep learning in recent years, which has shown a strong representation ability in image, text, and voice data processing \cite{Krizhevsky2012ImageNet, Socher2013Recursive, Hinton2012Deep}. The multi-layer architecture including non-linear functions designed in our model can capture the complex and non-linear impacts of contexts on groups, users, and events.

In order to model the situation that the influence weights of contexts  may change with the type of events, inspired by the neural attention mechanism \cite{Bahdanau2015Neural, Xiao2017Attentional}, we design a neural attention network which learns the influence weights of contexts on users/groups from interactions among users/groups, contexts, and events instead of interactions just between users/groups and contexts.

In order to capture the group preference more accurately, we propose that the calculation of a group preference should include two aspects: the group indirect preference and the group direct preference.The former could be obtained by aggregating members' preferences by employing  an attention mechanism, which can learn group aggregation strategy adaptively from data, while the latter could be learned from the historical interactions between groups and events with neural networks.

    In summary, the contributions of this paper are listed as follows:

1)	We propose an Attention-based Context-aware Group Event Recommendation model (ACGER) in EBSNs. The model can effectively capture the complex and non-linear influence of contexts on users, groups, and events. As far as we know, this is the first work which addresses the context-aware group event recommendation from the perspective of neural representation learning.

2)	We design a novel neural attention mechanism, which not only models the interaction between users/groups and contexts, but also incorporates the impacts of events, so that the dynamic change of contextual weights with different events can be captured in time.

3)	We propose that the calculation of a group preference should not only consider the indirect preference obtained from group members, but also consider the direct preference which is completely different from the preference of each member. To aggregate member preferences to obtain the indirect preference, an adaptive group aggregation strategy based on a neural attention mechanism is proposed. And the group direct preference is learned from group-event interaction data by neural networks.

4)	Extensive experiments on two real datasets from Meetup show that the proposed model ACGER can achieve better recommendation performance compared with the state-of-the-art models.

The rest of this paper is organized as follows. Section \uppercase\expandafter{\romannumeral2} reviews the related works. Section \uppercase\expandafter{\romannumeral3} formulates our problem and presents the framework of our proposed model ACGER. In Section \uppercase\expandafter{\romannumeral4}, we elaborate the ACGER scheme which includes three main modules.
The experiments based on two real-world datasets are conducted and the performance analysis is given in Section \uppercase\expandafter{\romannumeral5}. The last section \uppercase\expandafter{\romannumeral6} concludes this paper and points out our future work.

\section{Related Work}
In this section, we review some works related to our problem
in the literature, including conventional context-aware recommendation methods, context-aware event recommendation for individuals, and context-aware event recommendation for groups.

\subsection{Conventional Context-Aware Recommendation Methods}
Context in the recommender system domain refers to any information that can be used to characterize the situation of an entity. An entity is a person, a place, or an object that is considered relevant to the interaction between a user and a recommender system \cite{Dey2001Understanding}.  Context-aware recommendation algorithms can be classified into three main algorithmic paradigms according to the phase when contextual information is incorporated: contextual pre-filtering, contextual post-filtering, and contextual modeling \cite{ricci2015recommender}.

 In contextual pre-filtering paradigm, contextual information is used for data selection or data construction. Then, ratings can be predicted using any traditional Two-Dimensional (2D) User$\times$Item recommender system on the selected data.  One early work is \cite{Adomavicius2005Incorporating}. It proposes a reduction-based approach, which reduces the problem of multidimensional (MD) contextual recommendations to the standard 2D recommendation space. In this work, the authors also tried to combine several contextual pre-filters into one model at the same time, which provides significant performance improvements over the one pre-filter approaches. \cite{Baltrunas2009Towards} proposes User Splitting technique, which splits the user profile into several sub-profiles, and each sub-profile represents the user in a particular context. \cite{Baltrunas2014Experimental} proposes Item Splitting technique, which splits each item into several fictitious items based on the contexts. \cite{Zheng2014Splitting} splits both users and items in the data set to boost context-aware recommendations.

In contextual post-filtering paradigm, contextual information is initially ignored, and any traditional 2D recommender system could be used on the entire data to predict the ratings. Then, the recommendation result is adjusted by using the contextual information. \cite{Panniello2009Experimental} introduces two contextual post-filtering methods: Weight and Filter. The Weight method adjust the recommendation list by reordering the recommended items according to their probability of relevance in the specific context, and the Filter method filters out recommended items that have low probability of relevance in the specific context. One important benefit of both contextual pre-filtering approaches and contextual post-filtering approaches is that all the previous research on 2D recommender systems could be directly applied. However, all these approaches require manual supervision and fine-tuning in the recommendation process.

In contextual modeling paradigm, contextual information is incorporated directly in the recommendation model as an explicit predictor of a user's rating for an item.  Some studies work on contextualize Matrix Factorization (FM) approach.  \cite{Baltrunas2011Matrix} presents Context-Aware Matrix Factorization (CAMF), which extends MF by considering the influence of contexts on items. \cite{Zheng2014CSLIM} extends a matrix factorization method SLIM (Sparse Linear Method) to a Contextual SLIM (CSLIM) incorporating contextual conditions for the top-$N$  recommendation task. However, these approaches cannot handle the ternary relational nature of data. Tensor Factorization (TF) is an extension of MF techniques to incorporate   different contexts as multifaceted user-item interactions in the recommendation process. One classical method is Multiverse Recommendation \cite{Karatzoglou2010Multiverse}, which relies on Tucker decomposition and allows to work with any categorical context. To address the implicit feedback, \cite{Shi2012TFMAP} proposed a ranking-based Tensor Factorization (TF) model by directly maximizing Mean Average Precision. Another significant work is proposed by \cite{Rendle2011Fast}. It applies Factorization Machines (MF) to model the interactions between each pair of entities in terms of their latent factors, such as user-user, user-item, user-context interactions.

As far as recommendation methods in EBSNs are concerned,  the existing works mostly adopt the paradigm of contextual modeling, which directly integrates context information into models to characterize the contextual influences.

\subsection{Context-Aware Event Recommendation for Individuals}
Many models have been proposed in the field of context-aware event recommendation for individuals. Qiao et al. \cite{qiao2014event} proposed a potential factor model to model online and offline social relations, geographical features of events, and implicit feedback of users in event social networks, so as to recommend offline events for users. Macedo et al. \cite{macedo2015context} extracted social relations, content, time, and geographical features respectively, and then used the learning to rank technology to combine these contextual information to generate event recommendation. Zhang et al. \cite{zhang2015collective} formulated the cold-start event recommendation problem, using Bayesian Poisson factorization as the basic unit to model different contextual factors, and further combined those units to form a unified model through a collective matrix factorization model. Xu et al. \cite{xu2019semantic} proposed a semantic-enhanced and context-aware hybrid collaborative filtering method, which combines semantic content analysis and contextual event influence for user neighborhood selection. Cao et al. \cite{Cao2018Multi} combined  multiple features about topology, temporal, spatial, and semantic to model user preferences, which alleviates the problem of data sparseness in EBSNs. With the successful application of deep learning and representation learning in the fields of image, speech and natural language processing in recent years \cite{Krizhevsky2012ImageNet, Socher2013Recursive, Hinton2012Deep}, some researchers have also applied these techniques in context-aware event recommendation. Wang et al. \cite{Wang2019Event2Vec} considered the temporal and spatial effects of  events, and mapped the event, location, and time into low-dimensional space based on event sequential data by representation learning method. Wang et al. \cite{Wang2018Deep} utilized convolutional neural network with word embedding to extract the high-level features of contextual information of a user's interested events and built up a user latent model for each user, then they incorporated the user latent models into a probabilistic matrix decomposition model to obtain more accurate recommendation performance. However, all these methods do not consider either the influence of contexts on group preferences nor the characteristics of group recommendation task (e.g., how to aggregate different preferences of members into a consistent group preference). Therefore, they cannot be directly applied to EBSN group recommendation.

\subsection{Context-Aware Event Recommendation for Groups}
Group recommendation in EBSNs has attracted more and more attention in recent years.Yuan et al. \cite{Yuan2014COM} proposed a probability model COM to simulate the generation process of group activities and  perform group recommendations. Purushotham et al. \cite{Purushotham2016Personalized} proposed a collaborative filtering algorithm based on the Bayesian model to recommend events for groups considering the potential topics of groups. Ji et al. \cite{Ji2018GIST} proposed a topic-based probability model for group recommendation, in which the group preference not only considers the interests of members, but also considers the interests of subgroups. Du et al. \cite{Du2019CVTM} proposed a probabilistic generative model to jointly learn groups' content preferences and venue preferences. They discovered a strong correlation between organizers and textual contents. Above methods focus on modeling the generative process of group preferences by utilizing the interaction among group members, lacking a deep investigation on the influence of contexts on group behaviors, resulting in a suboptimal performance for group recommendation. Recently, a method named GERF comprehensively considering the influence of various contexts on group recommendation has been proposed \cite{Du2020GERF}. It first models the influences of contexts including time, place, event content, and social relations on the user's preferences, then merges the preferences of users in a group to form the group preference. Finally, by using a learning to rank algorithm to learn the ranking function for each group, it produces the event recommendation lists for groups. This method relies on manually defined function to characterize the influences of contexts on users and events, which is insufficient to model the complex and highly non-linear influences of contexts. In addition, the influence weights of different contexts are not differentiated in this work. In fact, there are a few works recently, which could be used to model the influence of different contexts on users and events. Among them, \cite{He2017Neural} proposes Neural Factorization Machines (NFM) which enhances FM by modeling nonlinear feature interactions through neural networks. \cite{Xiao2017Attentional} proposes Attentional Factorization Machines (AFM) that improves FM by differentiating the importance of different feature interactions via a neural attention network. Similar to FM, these two models can be applied to the task of context-aware recommendations by specifying the input data. However, NFM fail to differentiate the different importance of context influences. AFM can automatically differentiate the importance of feature interactions, but it models the feature interactions in a linear way.  \cite{Mei2018Attentive} proposes a novel neural model named AIN to adaptively capture the interactions between contexts and users/items. And a neural attention mechanism is employed to model the influence weights of contexts on users/items. However, the neural attention mechanism of AIN neglects that the influence weight of a context on a user may change when the type of the item  concerned changes. In this paper, we employ the deep neural network and representation learning techniques to model the complex and non-linear interactions between contexts and entities including users, events, and groups, and propose a novel neural attention mechanism to weigh the influence of different contexts more accurately.

How to aggregate different members' preferences into a consistent group preference, i.e., the group aggregation strategy, has always been the focus of group recommendation research. In the early works, Masthoff \cite{Masthoff2004Group} proposed 10 aggregation strategies based on social choice theory, such as approval voting, Borda counting, least misery, average, etc. Ardissono et al. \cite{ardissono2003intrigue} assigned greater weight to people with special needs (such as children or disabled people). Berkovsky et al. \cite{berkovsky2010group} judged a user's activeness based on the number of items he/she has rated, and assigned greater weight to more active users. SEO et al. \cite{Seo2018Enhanced} considered the deviation of group members' opinions combining with the average and voting counting strategies. \cite{Quintarelli2016Recommending} measured the influence weight of a member by the number of times that his/her preference being consistent with the group's preference. However, the aggregation strategies in these methods are all predefined, which is data independent and lacks flexibility. When the decision-making strategy of a group change, the predefined aggregation strategy cannot adapt to it. To overcome above limitation, we propose an adaptive group aggregation strategy based on the neural attention mechanism, which can learn a member's weight from the data. In addition, the acquisition of a group preference in our method not only considers preferences of members, but also considers the preference that is quite different from each member's, which further improves the performance of group recommendation.
\section{Problem Formulation And Model Framework}
\subsection{Notations and Problem Formulation}
We use bold capital letters (e.g., $\boldsymbol X$) ) and bold lowercase letters (e.g., $\boldsymbol x$) to represent matrices and vectors, respectively. We employ non-bold letters (e.g. $x$) to denote scalars, and squiggle letters (e.g. ${\cal X}$) to denote sets.  $|\cdot|$ denotes the cardinality of a set. If not clarified, all vectors are in column forms.

 Given context variable ${\cal C}$ consisted of $k$ contextual factors, i.e.,  ${\cal C}=\{{{\cal C}_{1}},{{\cal C}_{2}},\ldots ,{{\cal C}_{k}}\}$, we suppose each contextual factor ${{\cal C}_{i}}$ has $|{{\cal C}_{i}}|$ values, denoted as ${{\cal C}_{i}}=\{c_{1}^{(i)},c_{2}^{(i)},\ldots ,c_{|{{\cal C}_{i}}|}^{(i)}\}$, where $c_{j}^{(i)}\text{ }(j\in \{1,\ldots ,|{{\cal C}_{i}}|\})$ is the $j$th value of the $i$th contextual factor ${{\cal C}_{i}}$, ${\cal L}$ is the set of tuple of values of contextual factors, denoted as ${\cal L}=\{({{c}^{(1)}}, c_{{}}^{(2)},\ldots ,{{c}^{(k)}})|{{c}^{(j)}}\in {{\cal C}_{j}}, j\in \{1,\ldots ,k\} \}$. Suppose we have a set of $n$ users ${\cal U}=\{{{u}_{1}},{{u}_{2}},\ldots ,{{u}_{n}}\}$, a set of  $m$ events ${\cal E}=\{{{e}_{1}},{{e}_{2}},\ldots ,{{e}_{m}}\}$, and a set of $s$ groups ${\cal G}=\{{{g}_{1}},{{g}_{2}},\ldots ,{{g}_{s}}\}$, where each group is consisted of a certain number of users, and we obtain the group-event interaction matrix $Y={{[{{y}_{ij}}]}_{s\times m}}$ and the user-event interaction matrix $R={{[{{r}_{ij}}]}_{n\times m}}$. Given a target group $g \in {\cal G}$, a tuple of values of contextual factors  $c=(c^{(1)},c^{(2)},\ldots ,c^{(k)})\in \cal L$ ,  our task is defined as recommending a list of events that group $g$ may be most interested in, i.e., top-$N$  event recommendation for group $g$.

\begin{figure*}[tbp]
	\centering
	\includegraphics[width=1\linewidth]{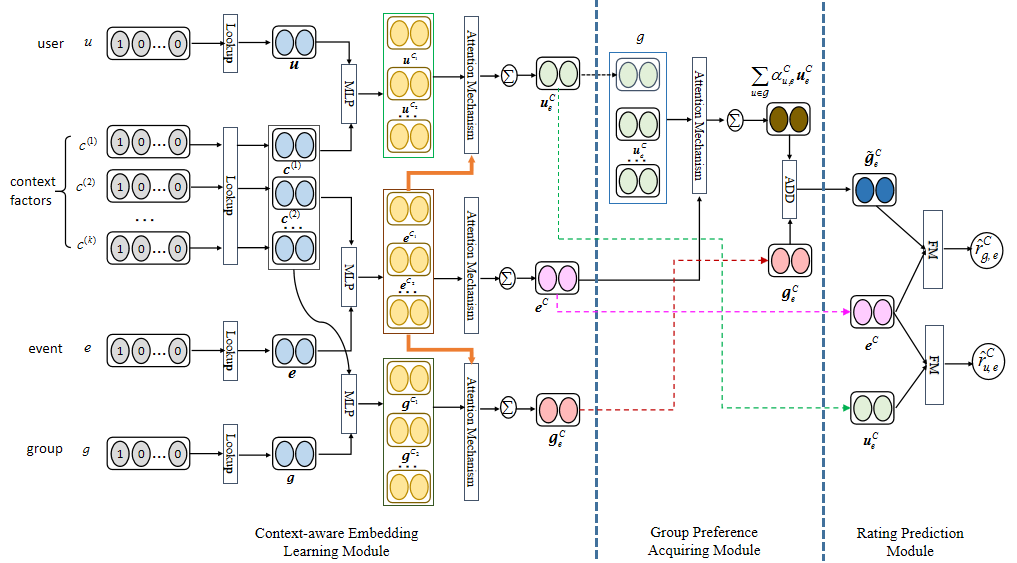}
	\caption{The framework of ACGER.}
	\label{fig:framework}
\end{figure*}
\subsection{Model Framework}
The ACGER model proposed in this paper is composed of three main components:

(1) Context-aware embedding learning module: given a group-event interaction record, the one-hot feature vectors of the related entities including group, event, group members, and contextual factors are taken as the initial inputs of the model, and they are mapped into low-dimensional and dense vectors through an embedding layer. Then, a Multi-Layer Perceptron (MLP) is used to capture the effects of interactions between contexts and users/events/groups, and then we get the enhanced embeddings of users/events/groups under the comprehensive influences of contexts through the neural attention mechanism. And the enhanced group embedding encodes the group's direct preference.

(2) Group preference acquisition module: we aggregate the group members' enhanced embeddings through a neural attention network to get the group's indirect preference. The indirect preference embedding and the direct preference embedding are combined to get the final group preference embedding.

(3) Score prediction module: the Factorization Machines (FM) model is used as the group score prediction layer, and a pairwise ranking loss is used for the model optimization.

Noted that in order to employ the abundant user-event interaction data in EBSNs to improve the accuracy of embedding learning, i.e., the learning of  the user embeddings and the event embeddings, we integrate the context-aware recommendation for individuals task into ACGER. Specifically, given a user-event interaction record, the one-hot feature vectors of user, event and contextual factors are fed into the model. Through an embedding layer, a multi-layer perceptron and a neural attention network in turn, we get the enhanced embeddings of the user/event under the comprehensive influences of the contextual factors. The FM model is still used to predict the score of a user to the target event. The overall framework of ACGER is shown in Fig. \ref{fig:framework}.

\section{Our Proposed ACGER Scheme}
\subsection{Context-aware Embedding Learning Module}
Given a set of contextual factors, the goal of this section is to obtain the feature representations of members, events and groups under the contextual influences. The module can be divided into three components and the detailed description is as follows:

1) Obtain the low-dimensional representations. Given contextual factors ${{\cal C}_1,{\cal C}_2, \ldots ,{\cal C}_k}$ and the tuple of their current value $c = ( {c^{(1)}},c_{}^{(2)}, \ldots ,{c^{(k)}}) $, the low-dimensional embedding representation of member $u$ is calculated as follows:
\begin{equation}
\boldsymbol u={{\boldsymbol P}^{T}} \tilde{\boldsymbol u},
\end{equation}
where $\boldsymbol P\in {{\mathbb{R}}^{|{\cal U}|\times d}}$ denotes the user embedding matrix,  $d$ denotes the dimension of the embedding vector, $T$ denotes the matrix transpose operation. And $\tilde{\boldsymbol u}\in {{\{0,1\}}^{\text{ }\!\!|\!\!\text{ }\cal U\text{ }\!\!|\!\!\text{ }}}$ is a one-hot vector, where the location of element 1 representing which row in the user matrix $\boldsymbol P$ the user $u$ corresponds to.

Similarly, we obtain the event $e$'s  embedding  $\boldsymbol e$ from the event embedding matrix $\boldsymbol Q\in {{\mathbb{R}}^{|{\cal E}|\times d}}$, and get the group $g$'s embedding $\boldsymbol g $ from  the group embedding matrix $\boldsymbol Z\in {{\mathbb{R}}^{|{\cal G}|\times d}}$. As for the  value $c^{(i)}$ of the contextual factor ${{\cal C}_{i}}$, its embedding $\boldsymbol c^{(i)}$ could be obtained from the $i$th context embedding matrix ${{\boldsymbol K}_{i}}\in {{\mathbb{R}}^{|{{\cal C}_{i}}|\times d}}$.

2) Obtain the embedding representation under the influence of each contextual factor. In order to capture the complex and non-linear impact of each contextual factor on members/events/groups, we use a MLP to map the input data to a deep, non-linear hidden space.

Specifically, we first concatenate the user embedding $\boldsymbol u$ with the context embedding ${{\boldsymbol c}^{(i)}}$, then we pass the concatenation through a stack of fully connected layers and finally get $u$'s influenced representation in the context of  ${{\cal C}_{i}}$. The formulation is as follows:
\begin{equation}\label{eq:mlp}
\begin{split}
 \boldsymbol f_{i,0}^{\psi }&= {\rm ReLU}(\boldsymbol W_{i,0}^{\psi }[\boldsymbol u,\boldsymbol c_{{}}^{(i)}]+\boldsymbol b_{i,0}^{\psi }),\\
 \boldsymbol f_{i,1}^{\psi }&={\rm ReLU}(\boldsymbol W_{i,1}^{\psi }\boldsymbol f_{i,0}^{\psi }+\boldsymbol b_{i,1}^{\psi }),\\
 &\cdots\cdots\\
 {{\boldsymbol u}^{{{\cal C}_{i}}}}&={\rm ReLU}(\boldsymbol W_{i,L}^{\psi }\boldsymbol f_{i,L-1}^{\psi }+\boldsymbol b_{i,L}^{\psi }),
 \end{split}
\end{equation}
where $[\cdot,\cdot ]$ denotes the concatenation of two vectors,  $\rm ReLU(\cdot )$ is the Rectifier activation function, $\boldsymbol W_{i,0}^{\psi }\in {{\mathbb{R}}^{d\times 2d}}$ and $\boldsymbol W_{i,j}^{\psi }\in {{\mathbb{R}}^{d\times d}}(j\in \{1,\ldots , L\})$ are the parameter matrices, $\boldsymbol b_{i,j}^{\psi }\in {{\mathbb{R}}^{d}}(j\in \{0,\cdots,{ L}\})$  denotes the bias vector, $\boldsymbol f_{i,j}^{\psi}\in {{\mathbb{R}}^{d}}(j\in \{0,\cdots, L-1\})$ denotes the output vector of the $j$th hidden layers. The superscript $\psi $ indicates the marked model parameters are related to Eq. \eqref{eq:mlp}.

As for event $e$ and group $g$, we can similarly get their contextual influenced representations ${{\boldsymbol e}^{{{\cal C}_{i}}}}$ and ${{\boldsymbol g}^{{{\cal C}_{i}}}}$  by passing the concatenations $[\boldsymbol e,\boldsymbol c_{{}}^{(i)}]$ and $[\boldsymbol  g,\boldsymbol c_{{}}^{(i)}]$ through  their respective MLP.

3) Obtain the unique embedding representation influenced by all contextual factors. Different context has different influence weight on members/events/groups. How to measure the weight accurately is a key problem. Inspired by the neural attention mechanism \cite{Bahdanau2015Neural, Xiao2017Attentional} which can learn the importance of  different components in the model from data, we consider using attention to learn the weights of various contextual factors.

In order to obtain the user's unique embedding representation under the influence of all contextual factors, we aggregate the user's embedding representations influenced by each context. To this end, it is necessary to calculate the influence weight of each context on the user. Different from the existing method \cite{Mei2018Attentive}, the influence weight of each context in our model not only considers the interaction between the user and the context, but also considers the event that the user is currently concerned about. Our idea is that we measure how much the user  $u$'s contextual representation ${{\boldsymbol  u}^{{{\cal C}_{i}}}}$ matches the event $e$'s contextual representation ${{\boldsymbol  e}^{{{\cal C}_{i}}}}$. The more they match, the more the user prefers the context ${{\cal C}_{i}}$ and accordingly ${{\cal C}_{i}}$ would be given more weight. Specifically, in order to calculate the attention score $\eta _{u,e}^{{{\cal C}_{i}}}$ between context ${{\cal C}_{i}}$ and user $u$ when $u$ is faced with event $e$, we design a neural attention network as follows:
\begin{equation}\label{eq:ucAtt}
\eta _{u,e}^{{{\cal C}_{i}}}={{\boldsymbol h}^{T}}{\rm ReLU}(\boldsymbol W_{1}^{^{\xi }}{{\boldsymbol u}^{{{\cal C}_{i}}}}+\boldsymbol W_{2}^{^{\xi }}{{\boldsymbol e}^{{{\cal C}_{i}}}}+{{\boldsymbol b}^{\xi }}),
\end{equation}
where $\boldsymbol W_{1}^{^{\xi }}\in {{\mathbb{R}}^{d\times d}}$ and  $\boldsymbol W_{2}^{^{\xi }}\in {{\mathbb{R}}^{d\times d}}$ are weight matrices of the attention network, ${{\boldsymbol b}^{\xi }}\in {{\mathbb{R}}^{d}}$  is the bias vector, $\boldsymbol h\in {{\mathbb{R}}^{d}}$ is a weight vector which projects the output of the ReLU activation function to a score value. The superscript $\psi $ indicates the marked model parameters are related to Eq. \eqref{eq:ucAtt}.

We normalize the value of $\eta _{u,e}^{{{\cal C}_{i}}}$ with a softmax function, and obtain the influence weight of context ${{\cal C}_{i}}$ on user $u$ when he/she is faced with $e$.
\begin{equation}
\beta _{u,e}^{{{\cal C}_{i}}}=softmax(\eta _{u,e}^{{{\cal C}_{i}}})=\frac{\exp \eta _{u,e}^{{{\cal C}_{i}}}}{\sum\limits_{i=1}^{k}{\exp \eta _{u,e}^{{{\cal C}_{i}}}}}.
\end{equation}

Finally we get $u$'s enhanced embedding influenced by all contextual factors, which is calculated as follows:
\begin{equation}
\boldsymbol u_{e}^{\cal C}=\sum\limits_{i=1}^{k}{\beta _{u,e}^{{{\cal C}_{i}}}}{{\boldsymbol u}^{{{\cal C}_{i}}}}.
\end{equation}

Similarly, the attention score $\mu _{g,e}^{{{\cal C}_{i}}}$ and the influence weight $\gamma _{g,e}^{{{\cal C}_{i}}}$ of context ${{\cal C}_{i}}$ on group $g$ when $g$ is faced with event $e$ are calculated respectively as follows:

\begin{gather}
\mu _{g,e}^{{{\cal C}_{i}}}={{\boldsymbol t}^{T}}{\rm ReLU}(\boldsymbol W_{1}^{\zeta }{{\boldsymbol g}^{{{\cal C}_{i}}}}+\boldsymbol W_{2}^{\zeta }{{\boldsymbol e}^{{{\cal C}_{i}}}}+{{\boldsymbol b}^{\zeta }})\label{eq:geAtt}\\
\gamma _{g,e}^{{{\cal C}_{i}}}=softmax(\mu _{g,e}^{{{\cal C}_{i}}})=\frac{\exp \mu
_{g,e}^{{{\cal C}_{i}}}}{\sum\limits_{i=1}^{k}{\exp \mu _{g,e}^{{{\cal C}_{i}}}}},
\end{gather}
where $\boldsymbol W_{1}^{\zeta }\in {{\mathbb{R}}^{d\times d}}$, $\boldsymbol W_{2}^{\zeta }\in {{\mathbb{R}}^{d\times d}}$, ${{\boldsymbol b}^{\xi }}\in {{\mathbb{R}}^{d}}$, $\boldsymbol t\in {{\mathbb{R}}^{d}}$ are model parameters. The superscript $\zeta$ indicates the marked model parameters are related to Eq. \eqref{eq:geAtt}.

Then, we get group $g$'s enhanced embedding influenced by all contextual factors, which  is calculated as follows:

\begin{equation}\label{eq:macro}
\boldsymbol  g_{e}^{\cal C}=\sum\limits_{i=1}^{k}{\gamma _{g,e}^{{{\cal C}_{i}}}}{{\boldsymbol g}^{{{\cal C}_{i}}}}.
\end{equation}
where $\boldsymbol  g_{e}^{\cal C}$  encodes  the group $g$'s direct preference.

At last, the attention score $\pi _{e}^{{{\cal C}_{i}}}$ and the influence weight  $\rho _{e}^{{{\cal C}_{i}}}$ of context ${{\cal C}_{i}}$ on event $e$   are calculated  as follows:
\begin{gather}
\pi _{e}^{{{\cal C}_{i}}}={{\boldsymbol s}^{T}}{\rm ReLU}(\boldsymbol W_{1}^{\varepsilon }\boldsymbol e+\boldsymbol W_{2}^{\varepsilon }{{\boldsymbol e}^{{{\cal C}_{i}}}}+{{\boldsymbol b}^{\varepsilon }}),\label{eq:eAtt}\\
\rho _{e}^{{{\cal C}_{i}}}=softmax(\pi _{e}^{{{\cal C}_{i}}})=\frac{\exp \pi _{e}^{{{\cal C}_{i}}}}{\sum\limits_{i=1}^{k}{\exp \pi _{e}^{{{\cal C}_{i}}}}},
\end{gather}
where $\boldsymbol W_{1}^{\varepsilon } \in {{\mathbb{R}}^{d\times d}}$, $\boldsymbol W_{2}^{\varepsilon } \in {{\mathbb{R}}^{d\times d}}$, ${{\boldsymbol b}^{\varepsilon }}\in {{\mathbb{R}}^{d}}$, and $\boldsymbol s\in {{\mathbb{R}}^{d}}$ are model parameters. The superscript $\varepsilon $ indicates the marked model parameters are related to Eq. \eqref{eq:eAtt}.

The event $e$'s enhanced embedding influenced by all contextual factors is calculated  as follows:
\begin{equation}
\boldsymbol e_{{}}^{\cal C}=\sum\limits_{i=1}^{k}{\rho _{e}^{{{\cal C}_{i}}}}{{\boldsymbol e}^{{{\cal C}_{i}}}}.
\end{equation}

\subsection{Group Preference Acquiring Module}
The goal of this section is to obtain an embedding vector for each group. In this paper, group preference is defined as the combination of indirect preference and direct preference. The group's direct preference is encoded in the group's enhanced embedding obtained by  Eq. \eqref{eq:macro}, and the group's indirect preference is obtained by aggregating the members' embeddings, in which the key problem is how to measure the influence weights of group members. We use the neural attention mechanism to learn them from data. Next, we elaborate on the process of obtaining group indirect preference.

Suppose group $g$ is making a decision on event $e$,  $\alpha _{u,e}^{\cal C}$ denotes the influence weight of member $u$ on the group $g$'s decision in the contexts $\cal C$, $\boldsymbol u_{e}^{\cal C}$ denotes $u$'s enhanced embedding influenced by all contextual factors, and embedding $\boldsymbol e_{{}}^{\cal C}$ denotes the property of event $e$ influenced by all contextual factors. Then $\alpha _{u,e}^{\cal C}$ is defined as the output of a neural attention network with embeddings $\boldsymbol u_{e}^{\cal C}$ and $\boldsymbol e_{{}}^{\cal C}$ as the inputs:
\begin{gather}
\delta _{u,e}^{\cal C}={{(\boldsymbol p)}^{T}}{\rm ReLU}(\boldsymbol W_{1}^{\tau }\boldsymbol u_{e}^{\cal C}+\boldsymbol W_{2}^{\tau }\boldsymbol e_{{}}^{\cal C}+{{\boldsymbol b}^{\tau }}),\label{eq:tao}\\
\alpha _{u,e}^{\cal C}=softmax(\delta _{u,e}^{\cal C})=\frac{\exp \delta _{u,e}^{\cal C}}{\sum\nolimits_{{u}'\in {{g}_{i}}}{\exp \delta _{{u}',e}^{\cal C}}},
\end{gather}
where $\delta _{u,e}^{\cal C}$ is the attention score between member $u$ and event $e$, $\boldsymbol W_{1}^{\tau }\in {{\mathbb{R}}^{d\times d}}$, $\boldsymbol W_{2}^{\tau }\in {{\mathbb{R}}^{d\times d}}$ are the parameter matrices of the attention network,  ${{\boldsymbol b}^{\tau }}\in {{\mathbb{R}}^{d}}$ is the bias vector, $\boldsymbol  p$ is a parameter vector projecting the value of ReLU function into a score. The softmax function normalizes the score value to get the final weight $\alpha _{u,e}^{\cal C}$. The superscript $\tau $ indicates the marked model parameters are related to Eq. \eqref{eq:tao}.

With the attention mechanism defined above, we can learn adaptively the aggregation strategy, which a group may change for different events, from interactions among contexts, groups and events. Next, we use the weights to aggregate the members' embeddings to form the group's indirect reference. In order to obtain the group's preference embedding, we combine the indirect preference with the direct preference by using an addition operation, which is utilized to combine different signals in the embedding space in work \cite{Xiao2017Attentional}). Specifically, when considering the event $e$, group $g$'s embedding, denoted as $\tilde{\boldsymbol g}_{e}^{\cal C}$, is calculated as follows:
\begin{equation}
 \tilde{\boldsymbol g}_{e}^{\cal C}=\sum\limits_{u\in {{\cal U}_{g}}}{\alpha _{u,e}^{\cal C}\boldsymbol u_{e}^{\cal C}}\text{ }+\text{  }\boldsymbol g_{e}^{\cal C},
\end{equation}
where $\sum\limits_{u\in {{\cal U}_{g}}}{\alpha _{u,e}^{\cal C}\boldsymbol u_{e}^{\cal C}}$ and $\boldsymbol g_{e}^{\cal C}$ denotes the indirect preference and the direct preference of group $g$ respectively,  ${{\cal U}_{g}}$ is the set of members of group $g$.
\subsection{Rating Prediction Module}
In order to predict the group rating, we select FM \cite{Rendle2010Factorization}) model. This is because the interaction data in EBSNs is very sparse, and FM can model the high-order interaction between features more effectively than other methods on the sparse dataset \cite{Guo2017DeepFM}. Specifically, we feed the concatenation of group embedding and event embedding  $\boldsymbol x=[\tilde{\boldsymbol g}_{e}^{\cal C},\boldsymbol e_{{}}^{\cal C}]$ into FM, then the predicted rating of $g$ on target event $e$ is calculated as follows:
\begin{equation}\label{eq:fm}
 \hat{r}_{g,e}^{\cal C}={{w}_{0}}+{{(\boldsymbol w_{1}^{\phi })}^{T}}\boldsymbol x+\frac{1}{2}\sum\limits_{k=1}^{p}{[{{({{(\boldsymbol v_{k}^{\phi })}^{T}}\boldsymbol x)}^{2}}-{{({{(\boldsymbol v_{k}^{\phi })}^{2}})}^{T}}{{\boldsymbol x}^{2}}]},
\end{equation}
where ${{w}_{0}}\in \mathbb{R}$ is the global bias,  $\boldsymbol w_{1}^{\phi }\in {{\mathbb{R}}^{2d}}$ is the parameter vector, $\boldsymbol v_{k}^{\phi }\in {{\mathbb{R}}^{2d}}$ is the  $k$th column vector of parameter matrix $\boldsymbol V\in {{\mathbb{R}}^{2d\times p}}$, the hyper-parameter $p\in \mathbb{N}_{0}^{+}$ denotes the  dimension of factorized parameters.  The superscript $\phi$ indicates the marked model parameters are related to Eq. \eqref{eq:fm}.

We rank the candidate events according to their predicted scores, and finally select the top-$N$ events to form the event list recommended for the group.

In addition to the group-event interaction data, there are also rich user-event interaction data in EBSNs. In order to reinforce the task of group event recommendation, we integrate the task of context-aware event recommendation for individuals into our model. Specifically, given a user-event interaction pair $(u,e)$ and the current values of contextual factors, the one-hot feature vectors of user $u$, event $e$, and values of contextual factors are taken as the initial input data, and are passed through the embedding layer, MLP, and the attention network in turn to get the enhanced user embedding  $\boldsymbol u_{e}^{\cal C}$ and the enhanced event embedding $\boldsymbol e_{{}}^{\cal C}$. Then, the concatenation of these two embeddings $\boldsymbol x=[\boldsymbol u_{e}^{\cal C}, \boldsymbol e_{{}}^{\cal C}]$ are fed into FM to get user $u$'s prediction score on target event $e$, denoted as $\hat{r}_{u,\text{ }e}^{\cal C}$. Since the two recommendation tasks share user embeddings, event embeddings and part of network weight parameters, the learning effect of group recommendation task is reinforced.

\subsection{Model Optimization}
We treat the group event recommendation task as a ranking task, and select the commonly used pairwise learning method BPR (Bayesian Personalized Ranking) \cite{rendle2009bpr} to optimize the model parameters.

The pairwise learning method assumes that the observed interaction events should have a higher recommended ranking than the unobserved interaction events. The optimization objective function of the group recommendation task is as follows:
\begin{equation}\label{eq:gloss}
 \sum\limits_{(g,e,{e}')\in {{\Re }_{train}}}{-\log \sigma ({{{\hat{r}}}_{g,e}^{\cal C}}-{{{\hat{r}}}_{g,{e}'}^{\cal C}})}+{{\lambda }_{\Theta }}{{\left\| \Theta  \right\|}^{2}}\,
\end{equation}
where $\sigma (\centerdot )$ denotes the logistic function, $\Theta $ is the parameters to be learned in the neural network, ${{\lambda }_{\Theta }}$ is the regularization hyper-parameter, ${{\Re }_{train}}$ is the training set in which $(g,e,{e}')$ denotes that group $g$ interacted with event $e$ and did not interact with event ${e}'$.

Similarly, the objective function of individual recommendation task is as follows:
\begin{equation}\label{eq:uloss}
\sum\limits_{(u,e,{e}')\in {{{{\Re }'}}_{train}}}{-\log \sigma ({{{\hat{r}}}_{u,e}^{\cal C}}-{{{\hat{r}}}_{u,{e}'}^{\cal C}})}+{{\lambda }_{{{\Theta }'}}}{{\left\| {{\Theta }'} \right\|}^{2}},
\end{equation}
where ${\Theta }'$ is the set of parameters to be learned in the neural network, ${{{\Re }'}_{train}}$ is the training set in which $(u,e,{e}')$ denotes that user $u$ interacted with event $e$ and did not interact with ${e}'$.

Stochastic gradient descent is used to minimize above objective functions.  The optimization algorithm for group recommendation task is summarized in Algorithm 1.
 And the recommendation algorithm for  ACGER model is presented in Algorithm. 2.
\begin{algorithm}
	\caption{Optimization algorithm for group recommendation task in ACGER}
	\LinesNumbered
	\KwIn {${{{\Re }}_{train}}$, learning rate $lr$, regularization hyper-parameter ${{\lambda }_{\Theta }}$, FM hyper-parameter $p$ .}
	\KwOut{updated model parameters $\Theta$.}%
    Initialize $lr$ and model parameters $\Theta$\;
	\Repeat{\emph{convergence}}{
       Draw $(g,e,{e}')$ from ${{{\Re }}_{train}}$\;
       Compute $\boldsymbol g$, $\boldsymbol e$, $\boldsymbol e'$,  $\boldsymbol c^{(i)}$ $(i\in \{1,\ldots ,k\})$, $\boldsymbol u$  $({\forall}u \in g)$ by equations similar to Eq. (1) //obtain the low-dimensional embeddings of entities\;
       Compute ${{\boldsymbol g}^{{{\cal C}_{i}}}}$, ${{\boldsymbol e}^{{{\cal C}_{i}}}}$, ${{\boldsymbol e'}^{{{\cal C}_{i}}}}$, ${{\boldsymbol u}^{{{\cal C}_{i}}}}$, $i\in \{1,\ldots ,k\}$ by equations similar to Eq. (2)//obtain  the embeddings under the influence of each context factor\;
       Compute $\boldsymbol  u_{e}^{\cal C}$, ${\forall}u \in g$,  by Eq. (3)-(5)\;
       Compute $\boldsymbol  g_{e}^{\cal C}$ by Eq. (6)-(8)\;
       Compute $\boldsymbol e_{{}}^{\cal C}$ by Eq. (9)-(11)\;
       Compute $g$'s embedding $\tilde{\boldsymbol g}_{e}^{\cal C}$ by Eq. (12)-(14) \;
       Compute $\hat{r}_{g,e,e'}^{\cal C}=\hat{r}_{g,e}^{\cal C}-\hat{r}_{g,e'}^{\cal C}$ by Eq. (15)\;
       \For {\emph{each parameter} $\theta$  \emph{in} $\Theta$}{
          $\theta \leftarrow \theta + {{l}_{r}} \cdot (\frac{{{e}^{-\hat{r}_{g,e,e'}^{C}}}}{1+{{e}^{-\hat{r}_{g,e,e'}^{C}}}}\cdot \frac{\partial \hat{r}_{g,e,e'}^{C}}{\partial \theta }+{{\lambda }_{\theta }}\cdot \theta )$;
       }
    }
    \textbf{return} $\Theta$.
\end{algorithm}
\begin{algorithm}
	\caption{Recommendation Algorithm for ACGER}
	\LinesNumbered
	\KwIn {${\cal U}$, ${\cal E}$, ${\cal C}$, ${\cal L}$, ${\cal G}$,  group-event interaction matrix $Y$, user-event interaction matrix $R$, target group ${g \in \cal G}$, candidate event set ${\cal E}_{cand}$, and given contextual values $(c^{(1)},c^{(2)},\ldots ,c^{(k)})$.}
	\KwOut{the recommended event list  $L_{g}$ for  group $g$.}%
    build  model ACGER \;
	initialize  model parameters\;
    \Repeat{\emph{convergence}}{

    model.training(${{{\Re }'}_{train}}$) based on Eq. (1) - (5), (9)-(11), (15),(17) //utilize user-event interactions\;

    model.training(${{{\Re }}_{train}}$) as Algorithm 1 //utilize group-event interactions\;
    model.evaluate(${{{\Re }'}_{test}}$) //evaluate the performance of individual recommendation\;
    model.evaluate(${{{\Re }}_{test}}$) //evaluate the performance of group recommendation\;
    }

	\For{$e$ in ${\cal E}_{cand}$}{
		 $\hat{r}_{g,e}^{\cal C}$=$model$.predit($g, e, (c^{(1)},c^{(2)},\ldots ,c^{(k)}) $) based on Eq. (1) - (15)\;
		}
    $ L_{g}=topN({\cal E}_{cand})$  //select $N$ events with the greatest predict scores of $\hat{r}_{g,e}^{\cal C}$ given contexts ${\cal C}$\;
	\textbf{return} $L_{g}$.
\end{algorithm}
\section{Performance Analysis}
In this section, we conducted extensive experiments on real datasets to answer the following research questions:

(1) How does our proposed model ACGER perform compared with the state-of-the-art group recommendation model?

(2) How is the effectiveness of our designed attention network for learning the contextual influence weight?

(3) How is the effectiveness of our designed attention network for learning the group aggregation strategy?

(4) How do the three components of the model --- attentive context-aware embedding learning, group embedding learning, and individual recommendation task contribute to the performance of ACGER?

\subsection{Experimental Settings}

\subsubsection{Datasets}
The datasets of this paper come from Meetup.com$\footnote{http://www.meetup.com}$, a popular EBSN platform. Through this platform, users can create events online, reply on whether to participate in events, join in various online social groups, and participate in events offline. We use the API interface provided by Meetup to obtain the relevant experimental data in 12 months of 2016. We choose to recommend events within the city scope, and choose New York and San Diego city in USA for experiments since they have the largest number of events published in 2016. We generate groups consisted of 2 to 6 users who often participate in events together, and collect events participated by each group and collect events attended by each user. In order to eliminate noise data and ensure the reliability of experimental results, the selected users and groups are required to participate in at least 10 events. In consideration of the universality of groups with the size of 2 users, the selected events are required to have at least two participants. After the data pre-processing, the statistics of the two city datasets are shown in Table \ref{tab:Statistics}, where \#U-E denotes the number of user-event interactions and \#G-E denotes the number of group-event interactions
\begin{table}[htbp]
\centering
 \caption{\label{tab:Statistics}Statistics of the Meetup datasets.}
 \begin{tabular}{lccccl}
  \toprule
  \centering
  \makecell[c]{City}&\#Users&\#Events&\#Groups&\makecell[c]{\#U-E } & \makecell[c]{\#G-E } \\
  \midrule
 New York & 2,849 & 10,024& 2,727& 288,447 &  \makecell[c]{101,141} \\
 San Diego & 2,419 & 10,685& 1,992& 287,469 &  \makecell[c]{70,239} \\
  \bottomrule
 \end{tabular}
\end{table}

Four contextual factors were considered in our experiment: organizer, venue, time, and event content. The values of organizer and venue can be converted into one-hot vectors according to their integer ID. And since the value of time is continuous and the value of content is textual, we need to perform data pre-processing  on these two factors to get their initial vector representations.

For the event content denoted as ${{\cal C}_{cont}}$, we regard each content text as a document, and all event content documents constitute a corpus. We use the natural language processing technology CBOW (Continuous Bag-of-Words) \cite{Glorot2010Understandinga} to map each word in the corpus into a low-dimensional word vector. The vector  of the content of an event $e$ is calculated as follows:
\begin{equation}\label{eq:uloss}
\boldsymbol l_{e}^{{{\cal C}_{cont}}}=\frac{\sum\limits_{w\in {{\cal W}_{e}}}{{{\boldsymbol z}_{w}}}}{|{{\cal W}_{e}}|},
\end{equation}
where ${{\cal W}_{e}}$ denotes the set of words in the content text of event $e$, ${{\boldsymbol z}_{w}}\in {{\mathbb{R}}^{d}}$ denotes the vector of word $w$. The content embeddings of all events are combined to form a pre-trained embedding matrix, which is used to initialize the content embedding matrix in ACGER.

The context of time represents the start time of an event. In order to map a continuous timestamp into a discrete time slot, we adopt a weekday-hour pattern, such as ``2 (day of the week), 16:00-17:00 (hour of the day)". Then, we get at most $7 \times 24$  discrete time slots for the context of time. Next, the time of each event is mapped to a $7 \times 24$-dimensional one-hot vector according to its time slot.

\subsubsection{Evaluation Metrics}
For each dataset, we rank the group-event/user-event interactions according to the start time of events. Then, we take the first $80\%$ of ranked interactions as the training set, $10\%$ as the validation set, and the last $10\%$ as the test set. Validation set is used for tuning the hyper-parameters. In the test set, the interacted events of each group/user are regarded as the real interested events to evaluate the recommendation performance of algorithms. Since we use the pairwise loss as our objective function, positive samples are selected from the events that the group/user has interacted with, and negative samples are selected from the events that the group/user has not interacted with.

In order to evaluate the performance of top-$N$ recommendation methods, we adopt three widely used evaluation metrics: precision (P@N), recall (R@N), and NDCG (Normalized Discounted Cumulative Gain, NDCG@N)\cite{Yuan2014COM}. Among them, NDCG measures the ability of a method to rank the events of truly interest higher in the recommendation list. For each metric, the higher the value, the better the recommendation performance.
\subsubsection{Baselines}

To justify the effectiveness of our method, we compared it with the following methods:

(1)	GERF \cite{Du2020GERF}: This is a method of context-aware event recommendation for groups. In this method, context influences are defined manually, the group feature vector is obtained by concatenating the feature vectors of group members, and a simple linear model is adopted for predicting the group's rating for an event. This method does not differentiate the influences of different contextual factors and treats group members as equally important.

(2)	UL \cite{Seo2018Enhanced}: This is a traditional context-unaware group recommendation method. The group aggregation strategy in this method is manually predefined, which combines the deviation of group members' opinions with average and approval voting strategies.

(3)	AIN\_ACGER2 \cite{Mei2018Attentive}: This method uses AIN (Attention Interaction Network, [12]) to model the influence of contextual factors on users/groups and events. The difference between this method and ACGER is that this method doesn't consider the impact of events when calculating the influence weight of contextual factors on groups/users.

(4)	ACGER1\_LinerBpr: This is a variant of ACGER, which firstly obtains the embedding of users, events and groups under the influence of all contextual factors by using the same attention networks as that in ACGER. Then, the members' embeddings are concatenated to form the group embedding and a linear model is employed to predict the ratings (just the same as GERF). This method is used to compare with GERF. The difference between this method and GERF is that this method uses attention networks to model contextual influences rather than defining influences manually.
\subsubsection{Experimental Settings}

We implemented the neural network-based methods such as AIN\_ACGER2, ACGER1\_LinerBpr and our ACGER in PyTorch. Other methods such as GERF and UL are implemented in Python. For the methods based on neural networks, Adam algorithm is used for optimization. The minimum batch size and learning rate are respectively in the range of [128, 256, 512, 1024] and [0.001, 0.005, 0.01, 0.05, 0.1]. For the embedded layer and hidden layers, we use the Gaussian distribution with mean value of 0 and standard deviation of 0.1 to initialize their parameters randomly. In neural attention network, the embedding dimensions of users, groups, events and contexts are all empirically set to 32. For each MLP, we deploy it with two hidden layers with dimensions set to 48 and 40 respectively. The factorization dimension   in FM are set to 10. And the weight parameters of three elements in the calculation of predicted group rating in UL method are determined by grid search. We repeat 5 times for each setting to report the average results.
\subsection{Overall Performance Comparison (RQ1)}
\begin{figure*}[htbp]
\centering
\subfloat[New York]{
\label{fig:degreeCentralOn}
\begin{minipage}[t]{0.45\textwidth}
\includegraphics[width=1\linewidth]{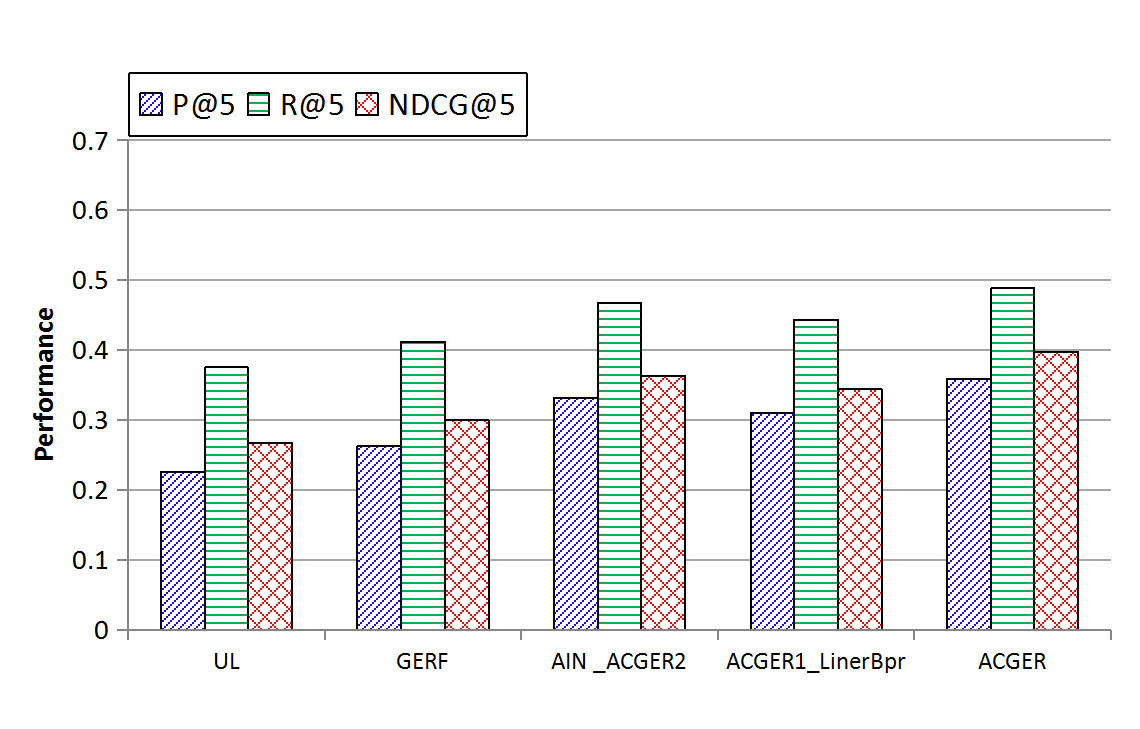}
\end{minipage}
}
\subfloat[San Diego]{
\label{fig:CloseCentralOn}
\begin{minipage}[t]{0.45\textwidth}
\includegraphics[width=1\linewidth]{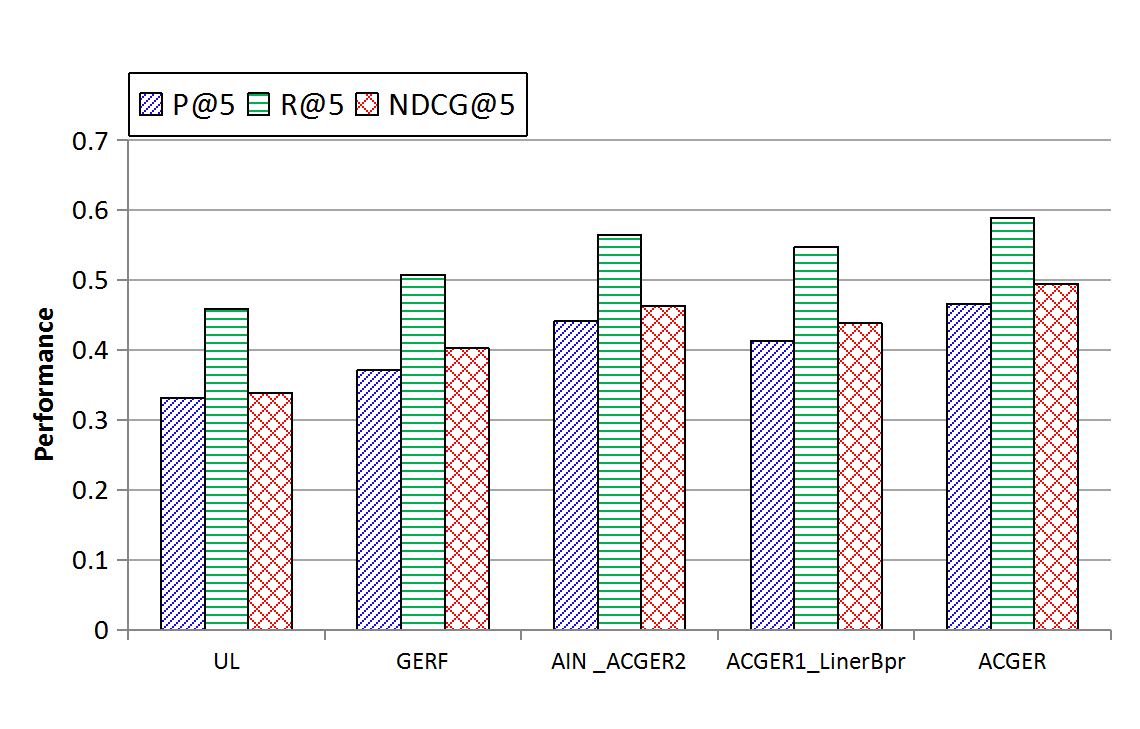}
\end{minipage}
}

\subfloat[New York]{
\label{fig:BetweenCentralOn}
\begin{minipage}[t]{0.45\textwidth}
\includegraphics[width=1\linewidth]{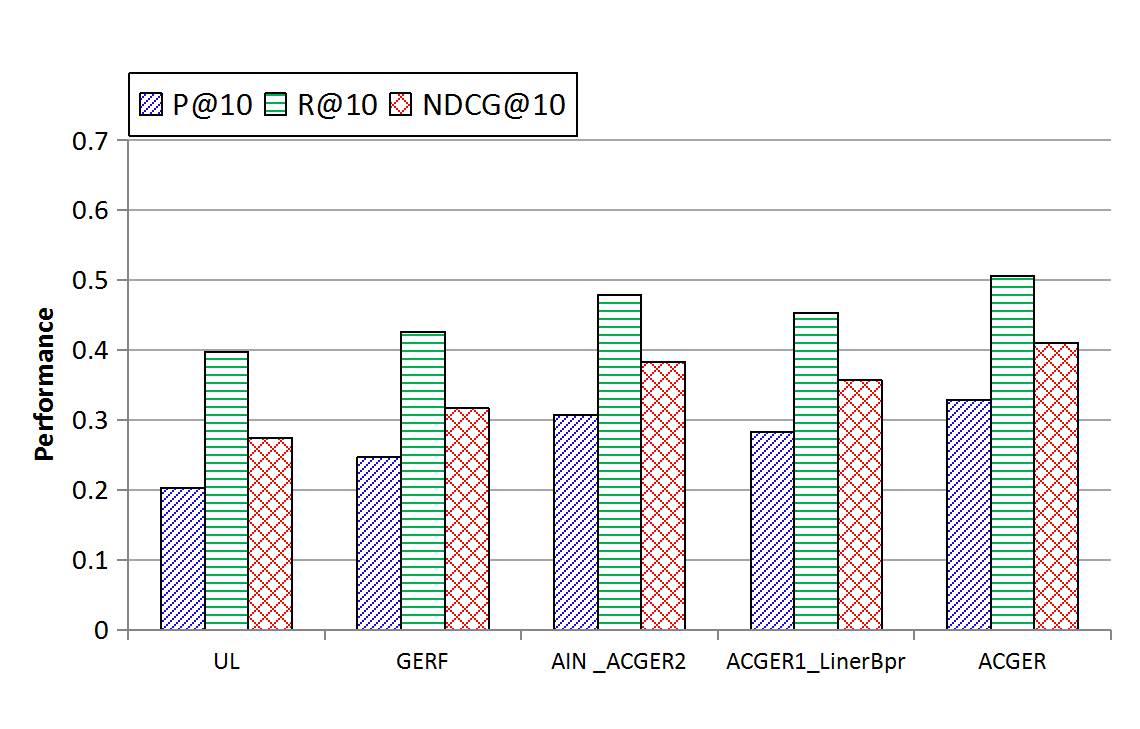}
\end{minipage}
}
\subfloat[San Diego]{
\label{fig:degreeCentralOn}
\begin{minipage}[t]{0.45\textwidth}
\includegraphics[width=1\linewidth]{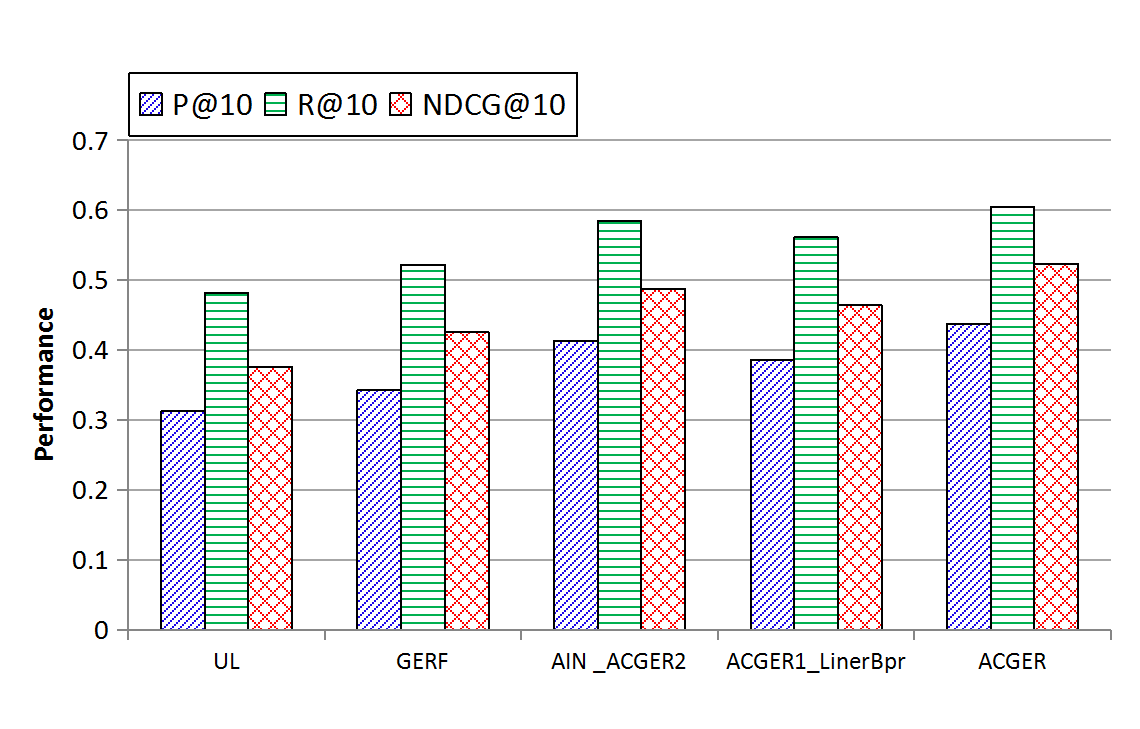}
\end{minipage}
}
\caption{Top-$N$ recommendation performance comparison between ACGER and baselines. }
\label{fig:baselines}
\end{figure*}

Fig. \ref{fig:baselines}  shows the top-$N$ ($N$=5, 10) recommendation performance of our ACGER and comparative methods on New York and San Diego datasets. We can see that ACGER achieves the best performance on both datasets with repect to three metrics. ACGER obtains improvements over the best baseline AIN\_ACGER2 by $2.8\%$ in P@5 , $2.1\%$ in R@5 and $3.4\%$ in NDCG@5 on New York dataset. On San Diego dataset, ACGER improves over  AIN\_ACGER2 by $2.5\%$ in P@5, $2.4\%$ in R@5 and $3.2\%$ in NDCG@5. This proves the validity of ACGER. Specifically, we can make the following observations: (1) context-aware methods (GERF, AIN\_ACGER2, ACGER1\_LinerBpr, and ACGER) have better performance than the context-unaware method (UL). This confirms the positive effect of context information on improving recommendation. (2) among the context-aware methods, the performance of neural network-based methods (AIN\_ACGER2, ACGER1\_LinerBpr, and ACGER) are better than that of method GERF which does not employ neural networks. This demonstrates the superiority of neural networks, especially their great ability in modeling the high-order interactions among different entities. (3) our model ACGER outperforms AIN\_ACGER2 on both datasets in three metrics. This is due to the fact that the influence of a contextual factor on a user/group may change when the type of an event concerned changes. Therefore, the performance can be further improved by taking into account the impacts of events when measuring the weight of contextual factors on users/groups. (4) The performance gap between ACGER1\_LinerBpr and ACGER shows the effectiveness of our model ACGER in modeling the group preference.

\subsection{Effect of Attention for Context Influence (RQ2)}
To demonstrate the effectiveness of attention mechanism of ACGER in distinguishing the influences of different contextual factors, we compare ACGER with its following variants:

1)	Avg\_ACGER2: In this method, the influence weights of different contextual factors on the entity (i.e., group, user, and event) are equal.

2)	AIN\_ACGER2: In this method, the influence weight of a contextual factor on users/groups is calculated by AIN method, in which the influence on users/groups only considers the interaction between users/groups and context, without capturing the dynamic change of the influence weight of the contextual factor on users/groups when the type of events changes.

3)	SingleU\_ACGER2: This method considers the influence of contextual factors on users and groups, neglecting their influence on events.

4)	SingleE\_ACGER2: This method considers the influences of contextual factors on events, neglecting their influence on users and groups.

\begin{figure*}[tbp]
\centering
\subfloat[New York]{
\label{fig:degreeCentralOn}
\begin{minipage}[t]{0.45\textwidth}
\includegraphics[width=1\linewidth]{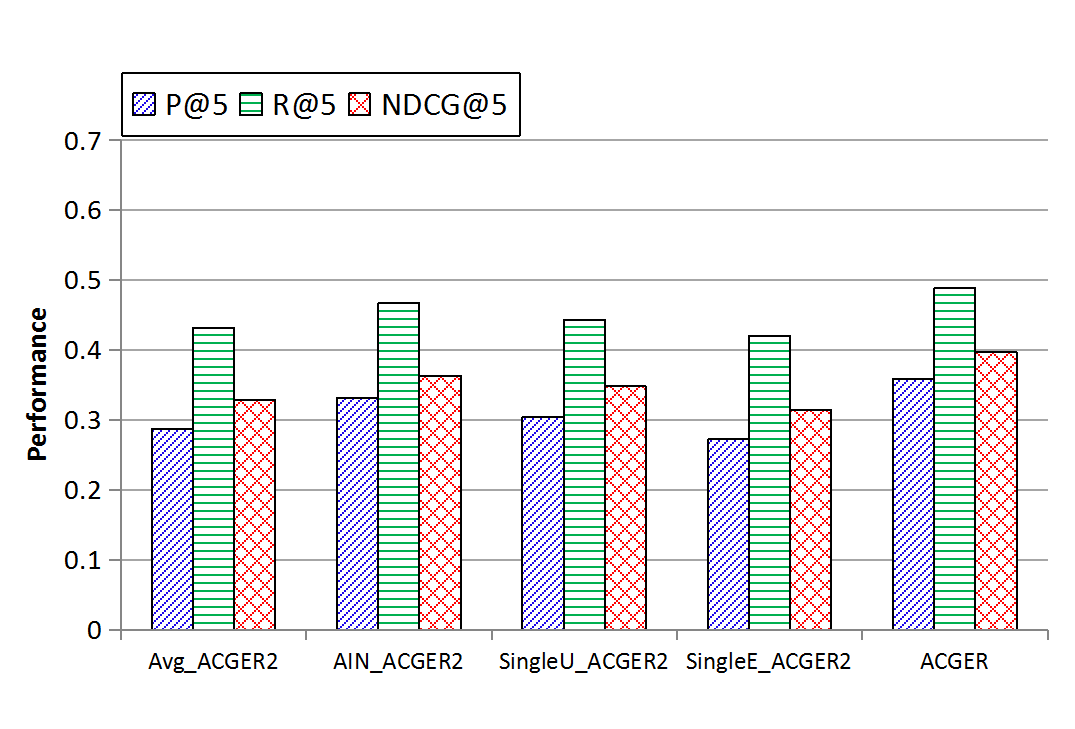}
\end{minipage}
}
\subfloat[San Diego]{
\label{fig:CloseCentralOn}
\begin{minipage}[t]{0.45\textwidth}
\includegraphics[width=1\linewidth]{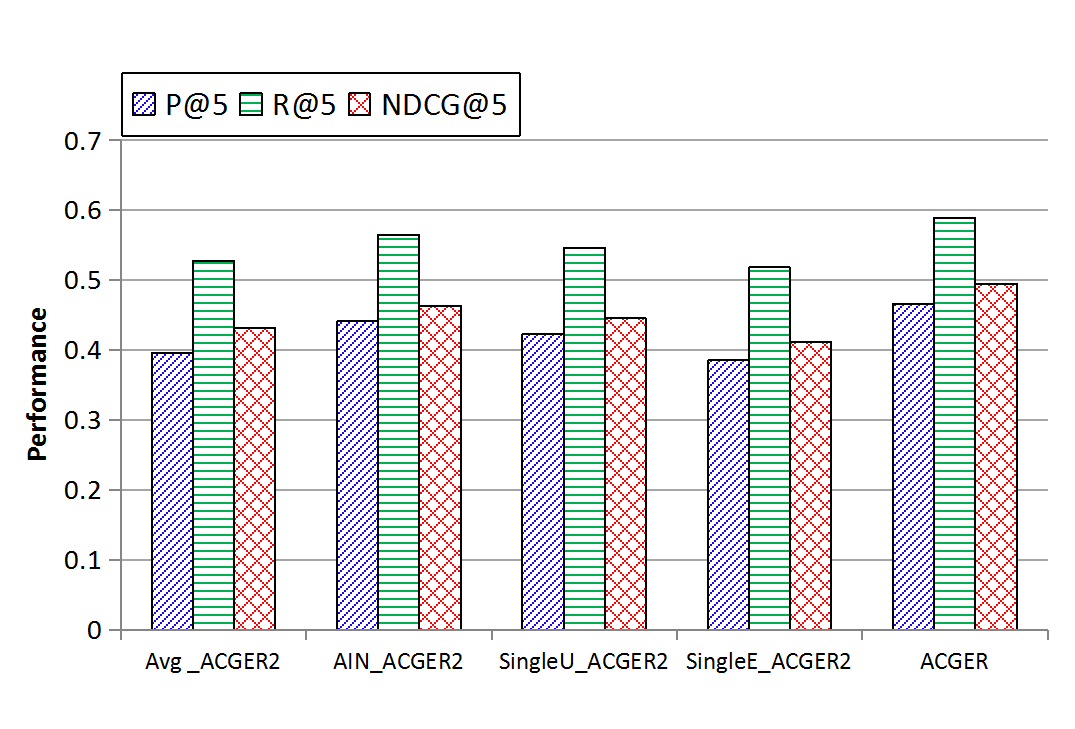}
\end{minipage}
}

\subfloat[New York]{
\label{fig:BetweenCentralOn}
\begin{minipage}[t]{0.45\textwidth}
\includegraphics[width=1\linewidth]{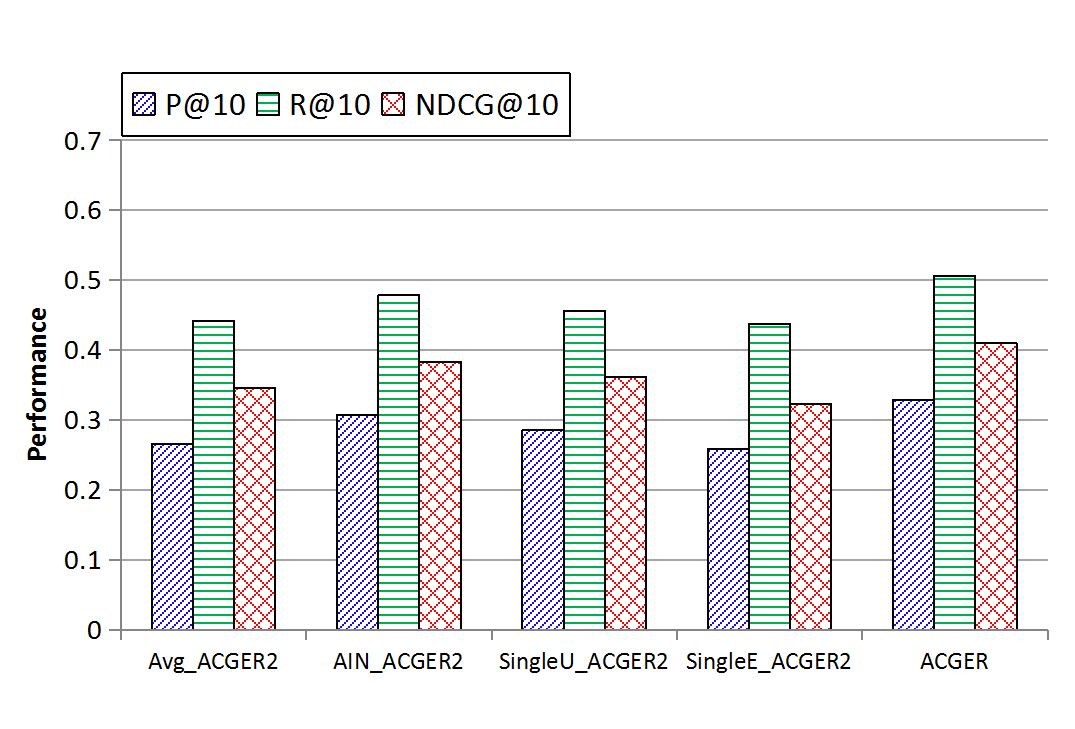}
\end{minipage}
}
\subfloat[San Diego]{
\label{fig:degreeCentralOn}
\begin{minipage}[t]{0.45\textwidth}
\includegraphics[width=1\linewidth]{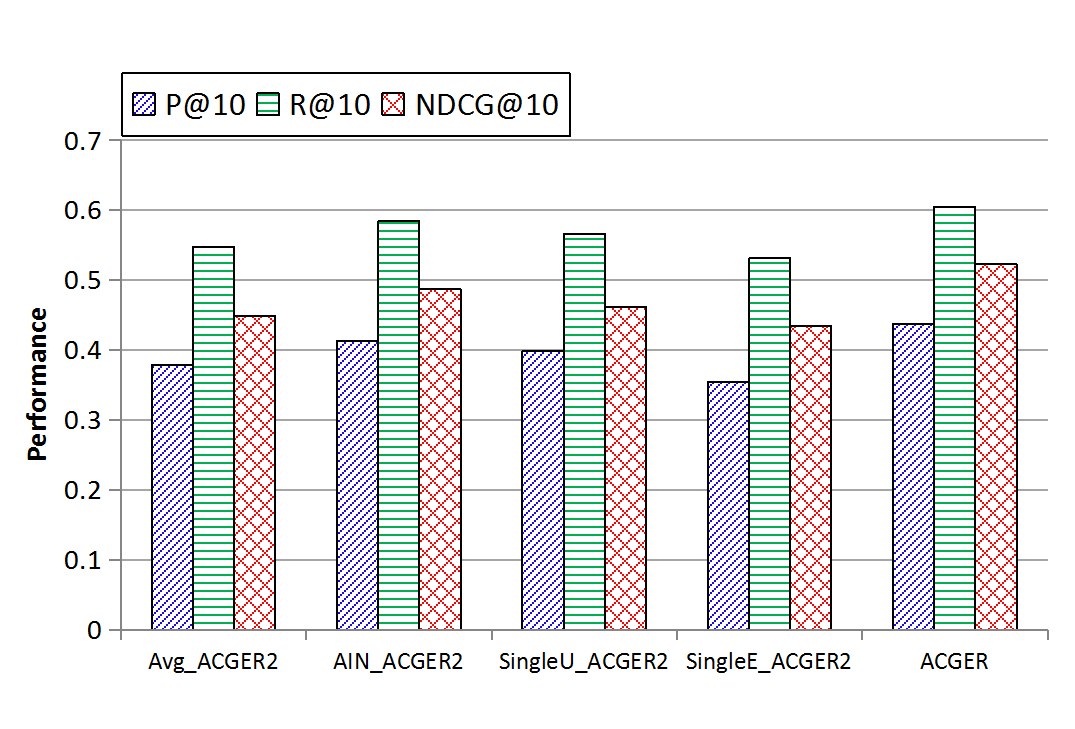}
\end{minipage}
}
\caption{Top-$N$ recommendation performance of ACGER and its variants in measuring the influence of contextual factors. }
\label{fig:ContextAtt}
\end{figure*}

Fig. \ref{fig:ContextAtt} shows the top-$N$ ($N$=5, 10) recommendation performance comparison results of ACGER and its variants on two datasets. We have the following observations: (1) The performances of Avg\_ACGER2 on both datasets are relatively low, slightly better than that of SingleE\_ACGER2, which performs the worst on both datasets. This is because that Avg\_ACGER2 does not distinguish the influence weights of different contextual factors. (2) the performance of AIN\_ACGER2 is not as good as that of ACGER, which indicates that modeling the dynamic change of influence weights of contextual factors on users/groups with the type of events helps to improve recommendation performance.  (3) ACGER consistently outperforms SingleU\_ACGER2 and SingleE\_ACGER2 on both datasets with respect to three metrics. This may be due to the fact that contextual factors characterize the situation where users/groups interact with events, and thus have influences on users/groups and events at the same time. Therefore, considering the effects of contexts on both users/groups and events leads to better performance. (4) SingleU\_ACGER2 slightly outperforms SingleE\_ACGER2 on both datasets, which indicates that considering the contextual influences on users/groups is more effective than considering the contextual influences on events to improve recommendation performance. It shows that the interests of users/groups are more sensitive to the contextual factors than the properties of events.

\subsection{Effect of Attention for Group Aggregation (RQ3)}
To demonstrate the effectiveness of the attention-based group aggregation strategy, we replace the strategy in ACGER with other predefined strategies, and obtain the following variants:

1)	ACGER1\_Avg \cite{berkovsky2010group}: The group aggregation strategy in this method adopts the average strategy.

2)	ACGER1\_BC \cite{Masthoff2004Group}: The group aggregation strategy in this method adopts the ``Borda Count (BC)" strategy, which scores ratings based on the ranking results. Specifically, for each group member, first rank the events according to their ratings, then score each event according to its ranking position (for example, the lowest ranking scores 0 and the highest ranking  $n-1$ in the $n$  events. Finally, scores of all members for each event are summed, and events are ranked according to the summed scores to get the recommendation list for the group.

3)	ACGER1\_Exp \cite{Quintarelli2016Recommending}: The group aggregation strategy in this method adopts weighted sum method, where the weight of a member is decided by the number of events he/she attends. Generally speaking, the more events a user has participated in, the more expert he/she may be.

4)	ACGER1\_MP \cite{Boratto2016Discovery}: The group aggregation strategy in this method adopts ``Most Pleasure (MP)" strategy, which takes the highest value in members' ratings as the group's rating for a candidate event.

\begin{figure*}[tbp]
\centering
\subfloat[New York]{
\label{fig:degreeCentralOn}
\begin{minipage}[t]{0.45\textwidth}
\includegraphics[width=1\linewidth]{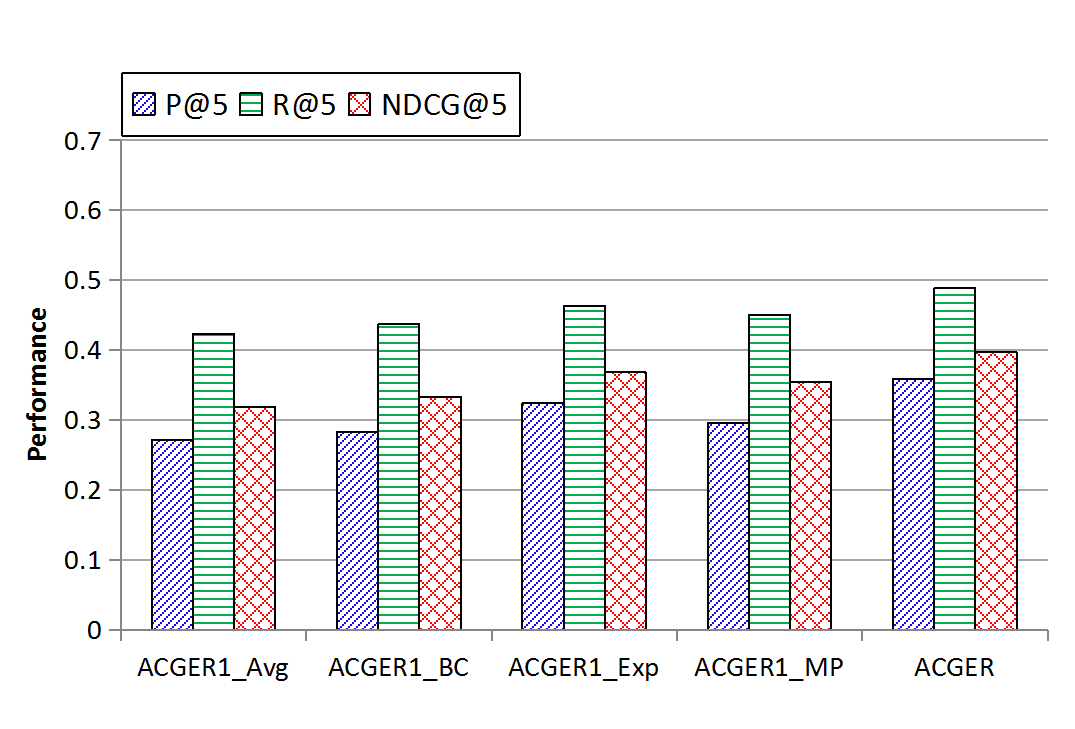}
\end{minipage}
}
\subfloat[New York]{
\label{fig:CloseCentralOn}
\begin{minipage}[t]{0.45\textwidth}
\includegraphics[width=1\linewidth]{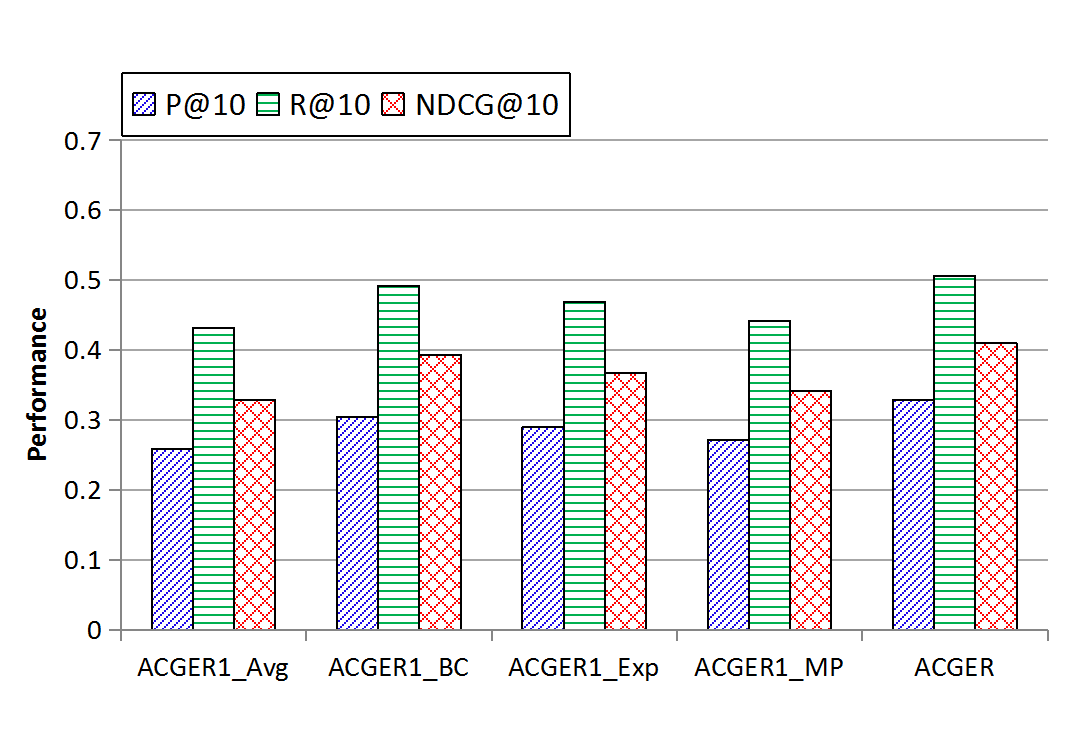}
\end{minipage}
}

\subfloat[San Diego]{
\label{fig:BetweenCentralOn}
\begin{minipage}[t]{0.45\textwidth}
\includegraphics[width=1\linewidth]{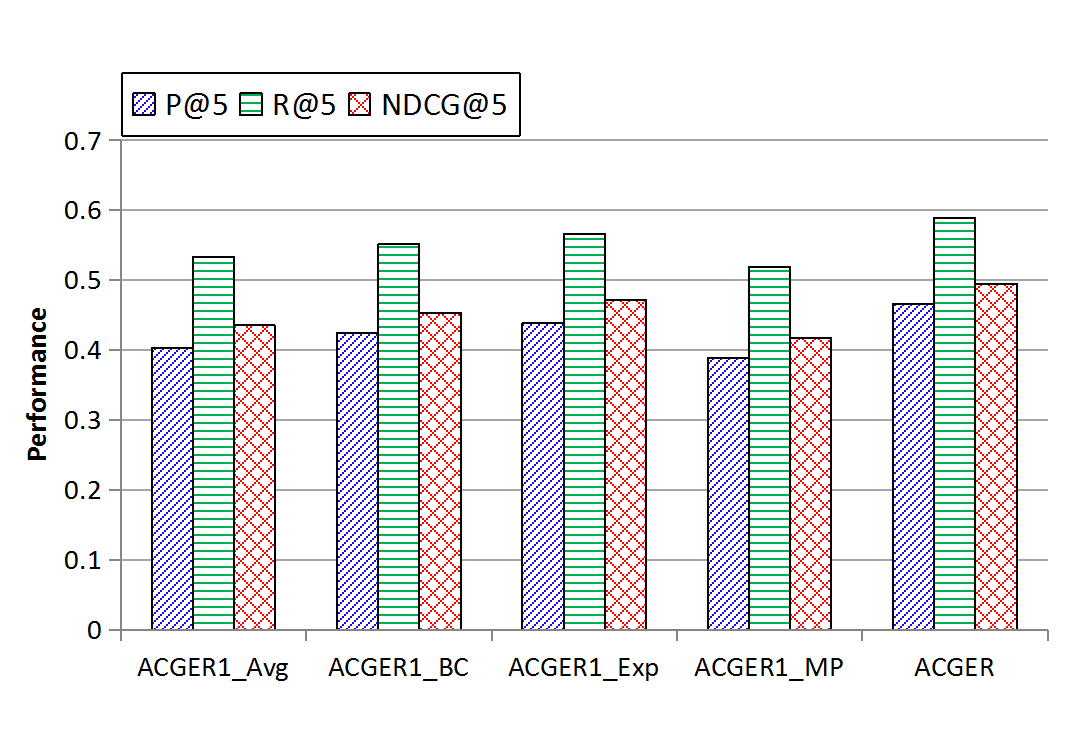}
\end{minipage}
}
\subfloat[San Diego]{
\label{fig:degreeCentralOn}
\begin{minipage}[t]{0.45\textwidth}
\includegraphics[width=1\linewidth]{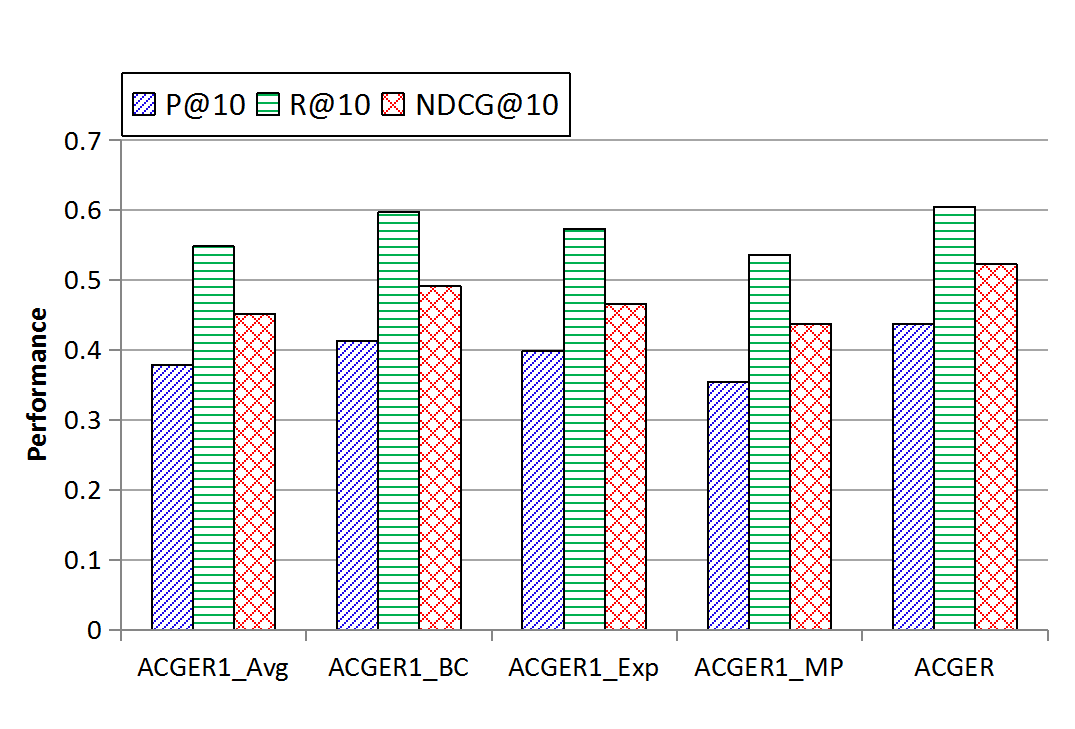}
\end{minipage}
}
\caption{Top-$N$ recommendation performance of ACGER and its variants in comparing group aggregation strategy. }
\label{fig:grpAtt}
\end{figure*}

The experimental results are shown in Fig. \ref{fig:grpAtt}. As you can see, there is no predefined strategy that always wins. For example, when $N$ = 5 on New York dataset, ACGER1\_MP ourperforms ACGER1\_BC (as shown in Fig. \ref{fig:grpAtt} (a)),  but underperforms when $N$ = 10 (as shown in Fig. \ref{fig:grpAtt} (b)). The Similar situation occurs on San Diego dataset. When $N$ = 5, ACGER1\_Exp outperforms ACGER1\_BC (as shown in Fig. \ref{fig:grpAtt} (c)), but underperforms when $N$ = 10 (as shown in Fig. \ref{fig:grpAtt} (d)).
%This demonstrates that the predefined strategies could not predict group preference well.
ACGER shows great flexibility and superiority because it can learn the group aggregation strategy from data.

\subsection{Contribution Analysis of Components (RQ4)}
\begin{figure*}[t]
\centering
\subfloat[New York]{
\label{fig:degreeCentralOn}
\begin{minipage}[t]{0.45\textwidth}
\includegraphics[width=1\linewidth]{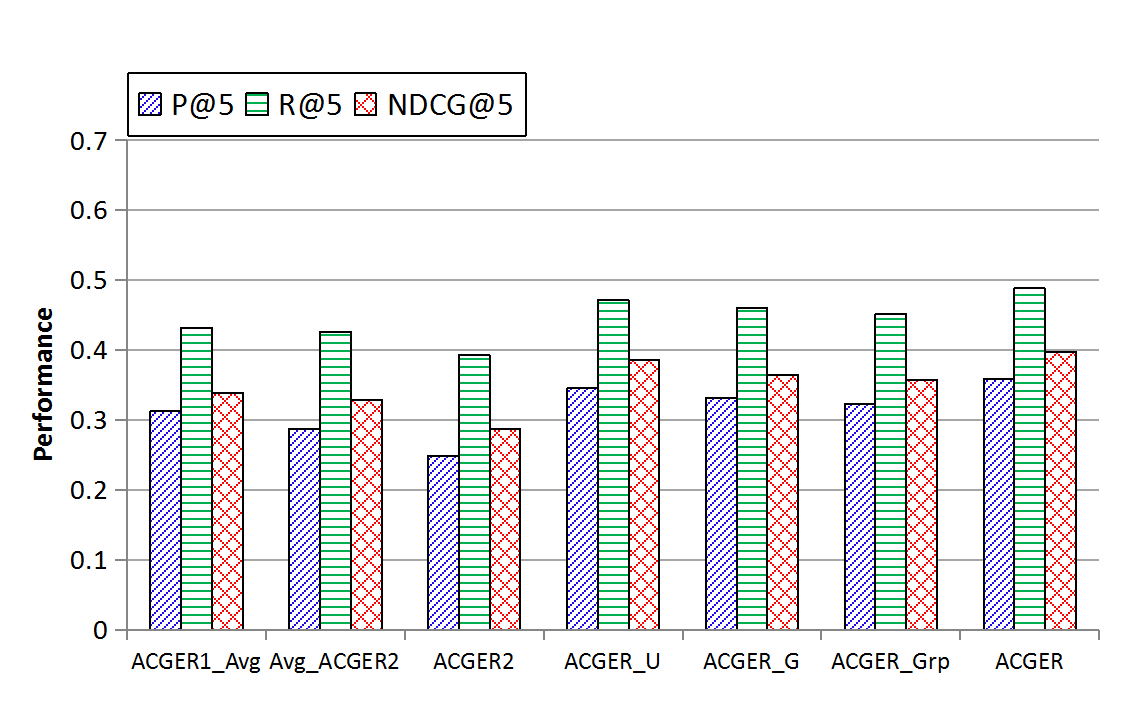}
\end{minipage}
}
\subfloat[San Diego]{
\label{fig:CloseCentralOn}
\begin{minipage}[t]{0.45\textwidth}
\includegraphics[width=1\linewidth]{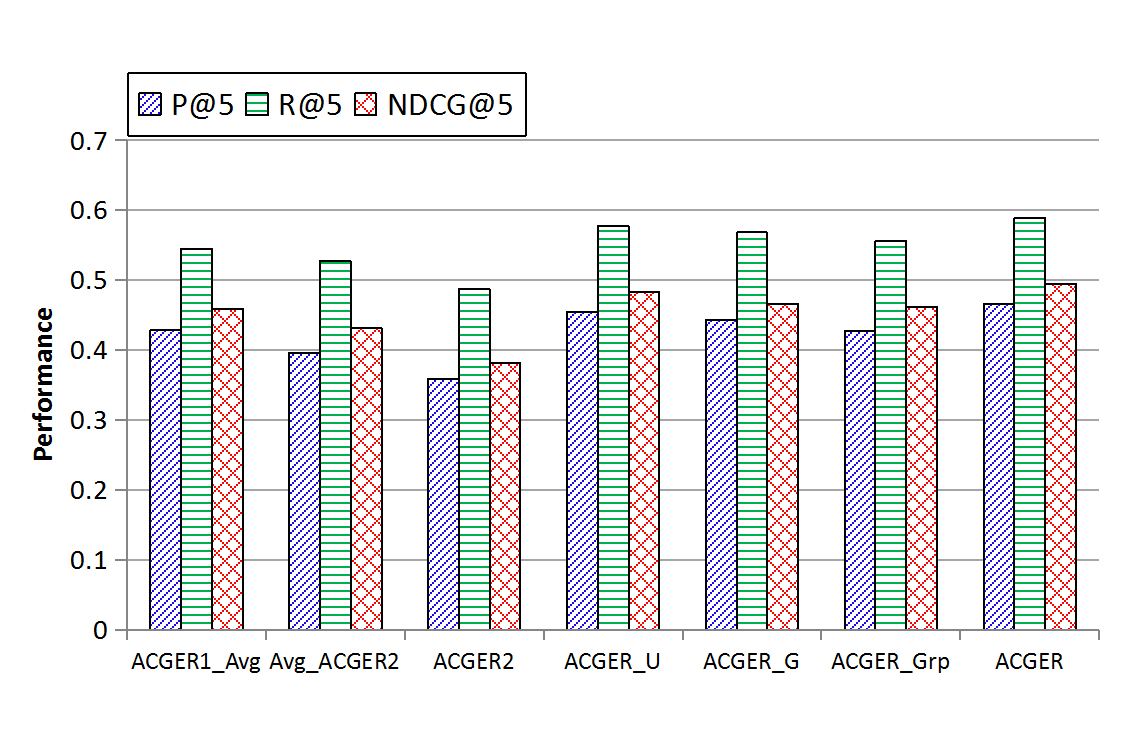}
\end{minipage}
}
\vspace{-1.8ex}
\subfloat[New York]{
\label{fig:BetweenCentralOn}
\begin{minipage}[t]{0.45\textwidth}
\includegraphics[width=1\linewidth]{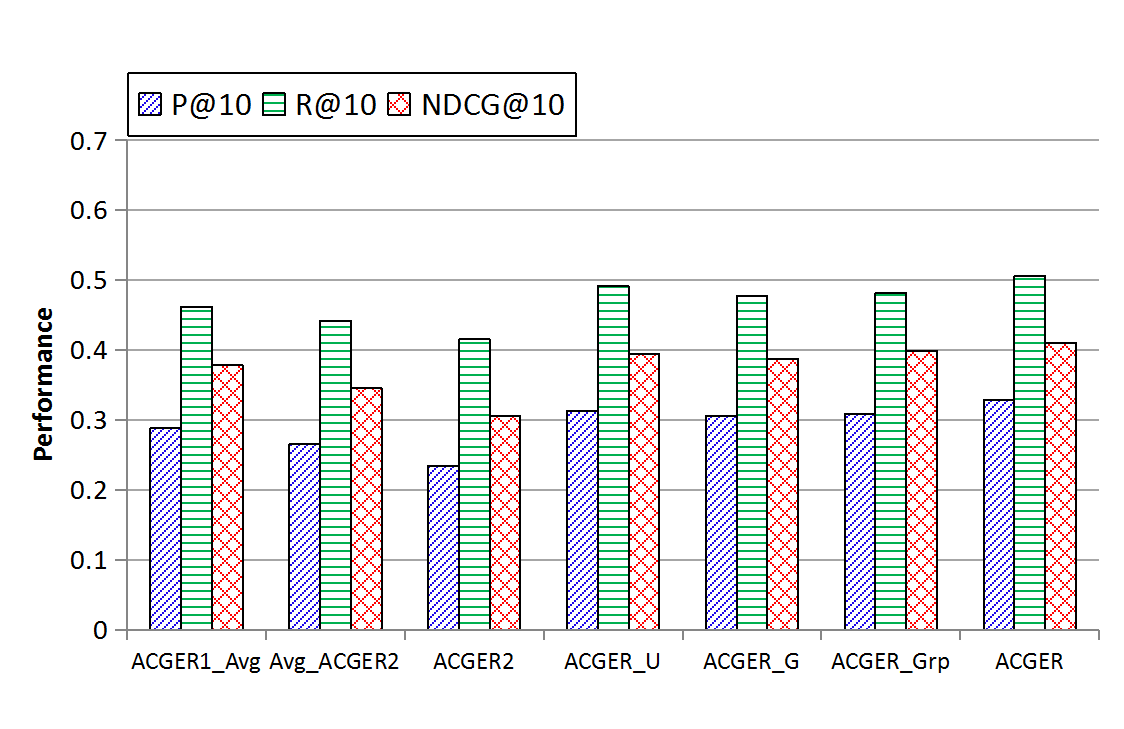}
\end{minipage}
}
\subfloat[San Diego]{
\label{fig:degreeCentralOn}
\begin{minipage}[t]{0.45\textwidth}
\includegraphics[width=1\linewidth]{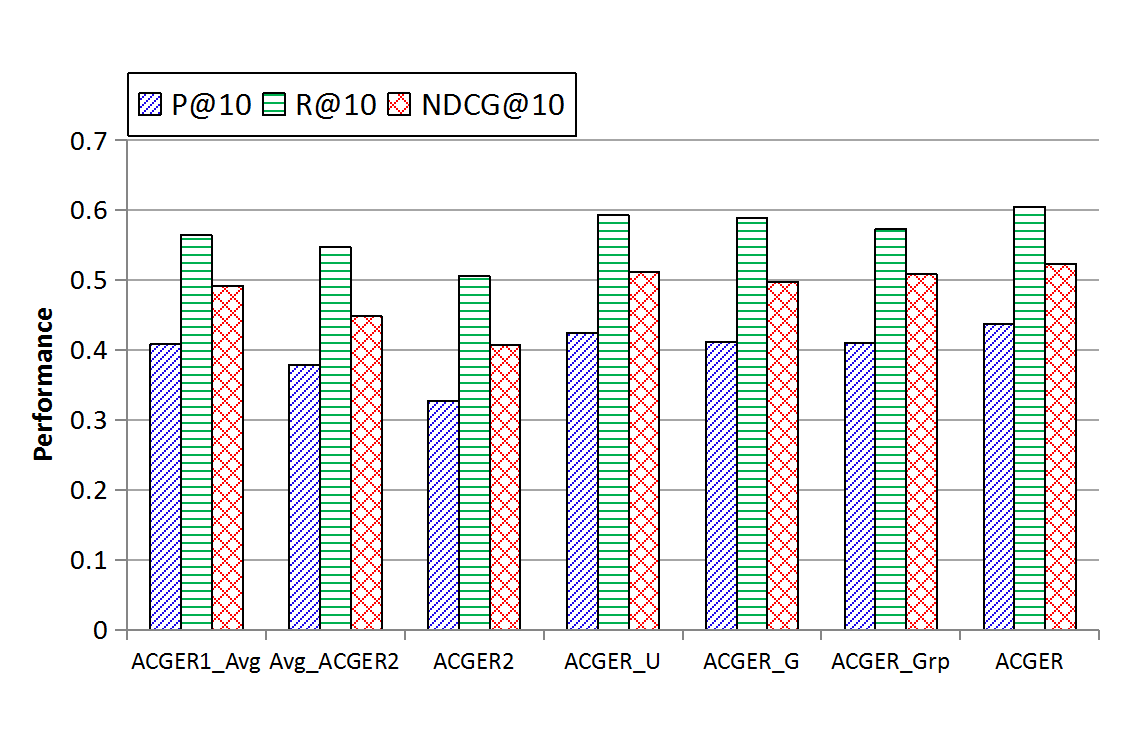}
\end{minipage}
}
\caption{Contribution analysis of different components of ACGER. }
\label{fig:componet}
\end{figure*}
To evaluate the contribution of the main components of ACGER to the group recommendation performance, we conducted some ablation experiments. We compare the ACGER with its following variants:

1)	Avg\_ACGER2, ACGER1\_Avg: Avg\_ACGER2 is the ACGER taking the contextual factors as equally important, and ACGER1\_Avg is the ACGER adopting the average strategy as the group aggregation strategy. The aim of these two variants is to study the contribution of attention networks.

2)	ACGER\_U, ACGER\_G: ACGER\_U is the ACGER with group indirect preference only, and ACGER\_G is the ACGER with group direct preference only. Our purpose is to study the contribution of two different type of group preferences to the recommendation performance.

3)	ACGER\_Grp: ACGER\_Grp is the ACGER without adding the individual recommendation task. The purpose of this method is to study the contribution of individual recommendation task to the group recommendation performance.

4)	ACGER2: This method denotes the ACGER without considering the influences of contextual factors. Our purpose is to study the contribution of modeling contextual influences to the recommendation performance.

The experimental results are shown in Fig. \ref{fig:componet}. We have observations as follows:  (1) Compared with ACGER, the performances of Avg\_ACGER2 and ACGER1\_Avg decrease on both datasets with respect to three metrics, which demonstrates that our attention-based weight calculation methods for contextual factors and group members are effective. (2) ACGER\_U and ACGER\_G underform ACGER on two datasets. This indicates that on two datasets, group embedding is affected by both group indirect preference and group direct preference. The performance of ACGER\_U is superior to that of ACGER\_G, which reveals that the group indirect preference has a larger impact in learning group preference on two datasets.  (3) the performance of ACGER\_Grp is inferior to that of ACGER on both datasets. For example, compared with ACGER, the performance of ACGER\_Grp decreases by $10.03\%$ in P@5, $7.58\%$ in R@5 and $10.08\%$ in NDCG@5 when $N$ = 5 on New York dataset, and the similar phenomenon could be observed on San Diego dataset. This indicates that the individual recommendation task could effectively reinforce the group recommendation task. (4) The performance of ACGER2 is significantly inferior to that of ACGER. This indicates that context information has a great impact on the performance of our model, because it can greatly improve the accuracy of group preference learning.

\subsection{Industrial Applications}
In the  market environment, participating in product exhibitions is one of the effective marketing methods for enterprises. Through the exhibition, the enterprise can show its products, enterprise strength, and brand image to the industry peers and live audiences, and it can quickly grasp the status and trend of domestic and international industries and get the information on the new products. By participating the exhibition,  the enterprise could find potential customers and new cooperative partners at a lower cost than general marketing channels. Therefore, enterprises have the inherent needs for participating in product exhibitions. An EBSN application provides a platform for organizers to release exhibition information and for enterprise users to sign up for the exhibitions. However, with the popularity of the platform, more and more information is published on the platform, and it becomes increasingly difficult for enterprise users, especially enterprise groups who often attend exhibitions together, to find their interested exhibitions. Therefore, it is urgent for the platform to provide exhibition recommendation service (i.e., event recommendation service) to help them find their interested exhibitions efficiently.

In addition to exhibitions, other types of events, such as technical seminars and professional summit forums, are also events of interest to enterprises. An EBSN event recommender system can also meet such needs.

\section{Conclusion And Future Work}
In this paper, we study an Attention-based Context-aware Group Event Recommendation model (ACGER)  for EBSNs. ACGER employs the neural networks with attention mechanism to model the complex and highly non-linear impacts of contexts on users, groups, and events. In order to model the situation where the influence weight of a contextual factor on users/groups may change with the type of events concerned, ACGER leverages a novel attention network, which not only considers the interaction between users/groups and context, but also considers the impacts of events. To overcome the limitation of lacking flexibility of the predefined group aggregation strategy in existing group recommendation methods, ACGER uses the neural attention mechanism to learn an adaptive group aggregation strategy from the data. Such mechanism enables a group automatically adjust its decision strategy according to the currently concerned events. Considering that a group often have different behavior patterns from its members, ACGER not only considers the indirect preferences aggregated from members' preferences, but also considers the direct preferences specific to group itself, which makes the group preferences captured more accurate. In addition, in order to make full use of user-event interaction data, we integrate the individual recommendation task into ACGER to reinforce the group recommendation task. Extensive experiments on two real datasets show that ACGER achieves higher recommendation performance than state-of-the-art methods.

In the future, as event sequence could be regarded as a special contextual factor, how to characterize the influence of event sequence on group recommendation could be further studied. We also try to explore the influence of trust relationship on user preference by using trust network embedding technology. In addition, an EBSN is a special heterogeneous network combining online and offline networks. Inspired by the recent development of graph neural network, how to employ this technique to model EBSN network structure for better recommendation is another interesting work.

\ifCLASSOPTIONcaptionsoff
  \newpage
\fi



%\bibliographystyle{IEEEtran}
%\begin{thebibliography}{1}

%\bibitem{zhang2015collective}

%\end{thebibliography}
\bibliography{ACGER_long}
%\end{IEEEbiography}


%\newpage



%\vfill


\end{document}